\documentclass[a4paper,11pt]{article}
\pdfoutput=1 

\usepackage{jcappub}

\usepackage[T1]{fontenc} 

\usepackage{hyperref}
\usepackage[dvipsnames]{xcolor}
\usepackage{subcaption}

\usepackage{float}

\usepackage[normalem]{ulem}

\usepackage{lineno}

\usepackage{cleveref}

\title{Probing Dark Matter Annihilation in the Galactic Centre with TRIDENT}

\author[a]{Yingwei~Wang$^{*}$,}
\author[a]{Xinhui~Chu$^{*}$,}
\author[a,b]{Andrew~Cheek$^{**}$,}
\author[a]{Iwan~Morton-Blake$^{**}$,}
\author[a]{Qichao~Chang,}
\author[a]{Gwenael Giacinti,}
\author[a]{Samy Kaci,}
\author[a,b,c]{Xin~Xiang$^{**}$,}
\author[a,b,c]{Donglian~Xu,}
\author[a]{ and Fuyudi~Zhang}

\affiliation[a]{State Key Laboratory of Dark Matter Physics, Tsung-Dao Lee Institute \& School of Physics and Astronomy, Shanghai Jiao Tong University, Shanghai 200240, China}
\affiliation[b]{Key Laboratory for Particle Astrophysics and Cosmology (MOE) \& Shanghai Key Laboratory for Particle Physics and Cosmology, Shanghai Jiao Tong University, Shanghai 200240, China}
\affiliation[c]{Hainan Research Institute, Shanghai Jiao Tong University, Hainan 572024, China}

\emailAdd{yingweiw@sjtu.edu.cn}
\emailAdd{xinh-chu@sjtu.edu.cn}
\emailAdd{acheek@sjtu.edu.cn}
\emailAdd{iblake@sjtu.edu.cn}
\emailAdd{xxiang@sjtu.edu.cn}

\footnotetext{$^{*}$These authors contributed equally to this work.}
\footnotetext{$^{**}$Corresponding authors.}

\abstract{We determine the future sensitivity of the TRIDENT neutrino telescope to dark matter annihilation in the Galactic Centre. By applying the full detector design and assuming an NFW halo profile, we show that TRIDENT will probe annihilation cross-section down to $\langle\sigma v\rangle\approx5\times10^{-27}\,{\rm cm}^3\,{\rm s}^{-1}$ for  $10\,{\rm TeV}$ dark matter, falling below the thermal freeze-out benchmark. The analysis is carried out with all-flavour neutrino interactions, where we demonstrate that cascade events, primarily due to $\nu_{e,\tau}$, show greater sensitivity to a dark matter signal compared to the more commonly studied track events. Furthermore, we highlight the impact of a previously overlooked background, Galactic neutrinos produced from interactions between hadronic cosmic rays and interstellar gas. We find dark matter sensitivities are more strongly degraded in the high energy region above $\sim 10\, {\rm TeV}$, with a maximal weakening of approximately a factor of  2. This effect remains smaller than the uncertainty associated with the dark matter density profile but can nonetheless mimic a positive annihilation signal. We contextualise these results with a concrete particle model and show that TRIDENT will be able to probe the most interesting untested parts of parameter space.}

\begin{document}
\maketitle
\flushbottom

\section{Introduction}

% WIMP and Thermal Freeze-out 
Unveiling the nature of dark matter (DM) remains a fundamental challenge in modern physics~\cite{Bertone:2016nfn, Cirelli:2024ssz}. Thermal freeze-out of heavy stable particles~\cite{Scherrer:1985zt, Gondolo:1990dk} provides an elegant production mechanism for DM in the early Universe while necessitating new interactions with the Standard Model (SM). This process is driven by the thermally averaged DM annihilation cross-section, $\langle \sigma v \rangle$. Moreover, for a large range of DM masses, $0.1-10^5\,{\rm GeV}$, the cross-section required to produce the correct relic varies very little and is approximately $\langle \sigma v \rangle \approx 3 \times 10^{-26} \, \text{cm}^3 \, \text{s}^{-1}$~\cite{Steigman:2012nb, Griest:1989wd}. Annihilation with such a cross-section should in principle be observable in regions of high DM density, motivating experimental searches. This method of \textit{indirect detection}~\cite{Gunn:1978gr, Stecker:1978du, Zeldovich:1980st} has made substantial progress over the past few decades, with the most successful experiments using photons and antiparticles as messengers~\cite{Hooper:2010mq, Fermi-LAT:2017opo, Fermi-LAT:2015att, AMS:2021nhj}. However, the thermal freeze-out target above $\sim 100\,{\rm GeV}$ remains largely untested~\cite{Leane:2018kjk}. At such high energies, while $\gamma$-ray telescopes have shown steady progress~\cite{HAWC:2017mfa,HESS:2018kom,CTA:2020qlo,LHAASO:2024upb}, neutrino telescopes have emerged as excellent astrophysical tools~\cite{Beacom:2006tt,Yuksel:2007ac,ANTARES:2015vis,Arguelles:2019ouk,Ishihara2023, KM3NeT:2024xca}. In this work we assess the future capabilities of the TRopIcal DEep-sea Neutrino Telescope (TRIDENT) precisely in the search for DM in the $10^3-10^5\,{\rm GeV}$ mass range.

The field of high-energy neutrino astronomy is entering a new era of rapidly expanding detector capabilities. Alongside the continued operation of IceCube, newly deployed facilities such as KM3NeT in the Mediterranean Sea~\cite{KM3NeT:2024xca} and Baikal-GVD in Lake Baikal~\cite{baikal2011} are now producing astrophysical neutrino measurements and steadily increasing the global sensitivity to high-energy neutrino sources. Looking further ahead, several next-generation observatories are under development, including IceCube-Gen2 at the South Pole~\cite{Ishihara2023}, the Pacific Ocean Neutrino Experiment (P-ONE)~\cite{P-ONE:2020ljt}, and TRIDENT in the South China Sea.

This paper presents the first model-independent sensitivity projections for TRIDENT’s DM annihilation search. By instrumenting a multi-cubic-kilometre volume of seawater with advanced optical sensors, it is expected to provide excellent sensitivity to both track- and cascade-like neutrino interactions, with  strong all-flavour performance at TeV--PeV energies~\cite{MortonBlake:2026allflavour}.
We target the Galactic Centre, demonstrating TRIDENT's potential for reaching cross-sections consistent with the DM thermal freeze-out benchmark. These projections guide pre-deployment detector optimisations and establish the analysis framework for future DM searches, positioning TRIDENT as a leading probe of the high-mass DM parameter space. 

An additional novel feature of this article is our treatment of the Galactic plane neutrinos, which are produced from interactions between hadronic cosmic rays and interstellar
gas. Such a background can significantly affect DM searches in the Galactic Centre and has not been included in previous studies~\cite{Miranda:2022kzs}. We adopt the Galactic plane neutrino flux inferred from the IceCube observations~\cite{icecubeneutrino} and quantitatively evaluate its impact on the TRIDENT sensitivity.

In this article, we first review the basics for determining the neutrino flux from DM annihilation in the centre of the Galaxy in Section~\ref{sec:dmflux}. In Section~\ref{sec:TRIDENT}, we describe the detector setup and simulation framework used to estimate the event rates for the DM signal and expected backgrounds. In Section~\ref{sec:limits} we present the TRIDENT sensitivity analysis and then evaluate the effects of the unknown Galactic neutrino background and halo profile uncertainties. In Section~\ref{sec:particle_model} we interpret these future projections for a specific particle physics model, exemplifying how TRIDENT will be able to uniquely probe important regions of the DM parameter space, and potentially discover it. Finally, we summarise our findings and discuss future directions in Section~\ref{sec:summary}. 

\section{Neutrino Flux From Dark Matter Annihilation}
\label{sec:dmflux}
For neutrino telescopes, the neutrino flux from DM annihilation is
\begin{equation}
\frac{d\Phi_\nu}{dE_\nu}=\frac{1}{4\pi}\frac{\langle \sigma v\rangle}{\kappa m_\chi^2}\left(\sum_\alpha \frac{dN^{\mathrm{prod}}_{\nu \alpha}}{dE_{\nu \alpha}}P_{\nu\alpha\rightarrow \nu\beta}\right) J(\Omega)  
\label{eq:flux}
\end{equation}
where $m_\chi$ is the DM mass and $dN^{\rm prod}_\nu/dE_\nu$ is the neutrino energy spectrum per annihilation, the superscript `${\rm prod}$' signifies that this spectrum is calculated at the production point. Typically, for the $\nu\bar{\nu}$ final state the energy spectrum is, to first approximation, simply a neutrino line, $dN_{\nu}/dE=2\,\delta(E_\nu -m_\chi)$. Similarly, the energy spectrum for the $4\nu$ final states can be approximated well by a box-shaped spectrum, $dN_\nu/dE=4H(m_\chi-E_\nu)/m_{\chi}$, where $H$ is the Heaviside step function~\cite{Ibarra:2012dw, Garcia-Cely:2016pse}. For energies above the electroweak scale, final state electroweak radiation can soften the spectra. Therefore, we use the spectra provided by HDMSpectra~\cite{Bauer:2020jay}. Unfortunately, the $4\nu$ final state is not included in the HDMSpectra, we estimate the electroweak correction to the box-shaped spectra by modelling it as a series of many neutrino lines and performing a weighted sum of corrected $2\nu$ spectra. For non-neutrino final states, such as quarks and charged leptons, neutrinos are still radiated from showering and particle decay, generating a continuous spectrum. For these we use the spectra provided in the code $\chi aro\nu$~\cite{Liu:2020ckq}\footnote{Note that we have checked our results are consistent when using other codes such as HDMSpectra~\cite{Bauer:2020jay} and CosmiXs~\cite{Arina:2023eic}.}. The propagation from the source to detection is encapsulated by the transition probability $P_{\nu_\alpha\to\nu_\beta}$. To calculate this, we take the best-fit PMNS matrix values from Ref.~\cite{nufit2025}. The numerical factor $\kappa$ is determined by the nature of DM. In the case of self-conjugate particles (such as Majorana fermions or real scalars), $\kappa=2$, whereas for non-self-conjugate particles (such as Dirac fermions or complex scalars), $\kappa=4$. We adopt $\kappa = 2$ for our model-independent analysis. Finally, $J(\Omega)$ in Eq.~(\ref{eq:flux}) is the integral of the DM density $\rho_\chi$ along the line of sight $dx$ (l.o.s.) over the target solid angle in the sky $d\Omega$, namely the J-factor
\begin{equation}
J\equiv \int d\Omega \int _{\mathrm{l.o.s.}} \rho_\chi^2(x)dx\,. 
\label{eq:J-factor}
\end{equation}

The precise form of the DM density profile is one of the main sources of uncertainty for our signal. For our model-independent bounds we will consider the benchmark NFW profile~\cite{Navarro:1995iw}  
\begin{equation}
\rho_\chi(r) = \frac{\rho_s}{(r/r_s)(1+r/r_s)^2} 
\end{equation}
where $r_s=20$ kpc is the scale radius and $\rho_s$ is the scale density, which is fixed by a local DM density $\rho_0=0.4$ GeV cm$^{-3}$ \cite{sofue2020rotation}. For each direction in the sky, we calculate the J-factor and use this as the spatial template for the DM signal. We discuss the dependence of our results with respect to the density profile choices in Section~\ref{subsec:dmhalo}.

\section{The TRIDENT Detector}
\label{sec:TRIDENT}

TRIDENT is a planned next-generation deep-sea neutrino telescope to be constructed in the South China Sea, at a depth of approximately 3.5 km. The full detector will comprise up to $\sim1000$ vertical strings, spanning several cubic kilometres, each instrumented with hybrid Digital Optical Modules (hDOMs) \cite{Shao:2025hDOM}.

\subsection{Neutrino Interactions in TRIDENT}

The TRIDENT detector is designed to achieve full-sky coverage and to detect all neutrino flavours with a high efficiency over a broad energy range, from sub-TeV to multi-PeV. Long optical scattering lengths in deep seawater allow for excellent photon timing and spatial resolution, boosting precision in the reconstruction of neutrino directions. These features make TRIDENT particularly well-suited for studying extended and diffuse astrophysical sources such as the Galactic Centre and for searching for DM annihilation signals.

Neutrino interactions in large-scale Cherenkov detectors such as TRIDENT produce two main event topologies: track and cascade-like events (referred to as tracks and cascades later in this work, respectively), corresponding to different neutrino flavours and interaction types.

\paragraph{Tracks} arise primarily from charged-current (CC) interactions of $\nu_\mu$, producing energetic muons that may traverse several kilometres in water. Their elongated light pattern enables precise directional reconstruction and provides the basis for point-source searches. The dominant background for these events comes from atmospheric muons generated by cosmic-ray interactions, which overwhelm the down-going event rate below $\sim100$ TeV. Concretely, we apply a cut of reconstructed tracks coming from below the horizon. 

\paragraph{Cascades} originate from CC interactions of $\nu_e$ and $\nu_\tau$, as well as neutral-current (NC) interactions of all flavours. They deposit energy within a more confined region (on the order of metres) compared to tracks, yielding a more limited angular resolution. However, for spatially extended or diffuse sources, such as the Galactic Centre, precise directionality is less critical, making cascades an important signal channel. Cascades also provide superior energy resolution, since their energy is typically fully contained within the detector, and suffer less contamination from atmospheric muons and neutrinos. Together with full-sky coverage and flavour inclusivity, these features make cascades an essential complement to tracks in DM searches. This study therefore considers both channels in combination, ensuring sensitivity to all neutrino flavours and interaction types.

\begin{figure}[h]
    \centering
    \includegraphics[width=\linewidth]{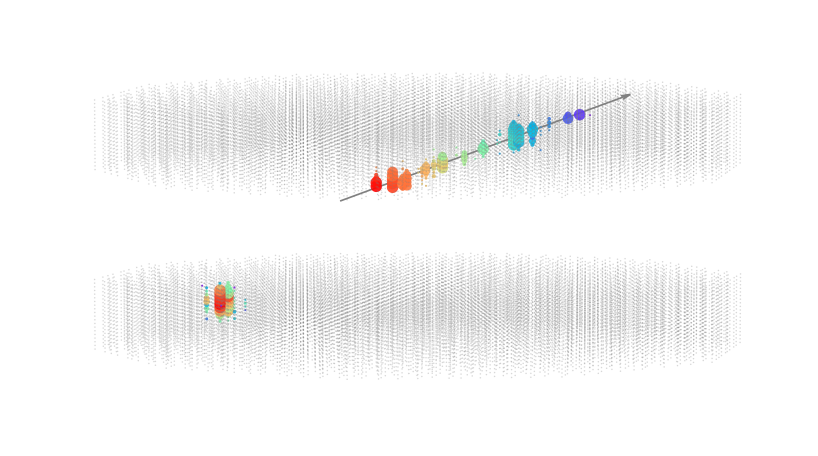}
    \caption{Visualisation of the two types of neutrino interactions from the TRIDENTSim framework~\cite{MortonBlake:2026allflavour}, showing an example $\nu_\mu$-CC track event (top) and $\nu_e$-CC cascade event (bottom). The colour of the spheres indicates the photon detection time, with red corresponding to early arrivals and blue to later arrivals, while the radii represent the number of photons detected by each hDOM.  }
    \label{fig:interactions_type}
\end{figure}

Example track and cascade events simulated using TRIDENT's simulation framework, described in the following section, are shown in Fig.~\ref{fig:interactions_type}, highlighting the significant differences in energy deposition possible for each interaction type within the detector.

\subsection{Simulation Framework}
\label{sec:simulation}

The detector response and sensitivity estimates presented in this work are based on the TRIDENTSim full-chain Monte Carlo framework~\cite{MortonBlake:2026allflavour}. The simulation models neutrino interactions of all flavours, including propagation through the Earth, deep inelastic scattering near the detector, and the subsequent production and propagation of secondary particles using \textsc{Pythia8}~\cite{pythia8}, \textsc{CRMC}~\cite{ulrich2021cosmic}, and \textsc{Geant4}~\cite{geant4}. Cherenkov photon production and propagation are simulated with \textsc{Geant4} and the \textsc{OptiX} ray-tracing engine~\cite{blyth2019opticks}, incorporating the measured optical properties of seawater at the TRIDENT site~\cite{Ye:2023trident}. The reconstruction of key neutrino direction and energy parameters is carried out using dedicated track and cascade likelihood-based algorithms~\cite{MortonBlake:2026allflavour}. These define the effective areas and angular resolutions used in the DM sensitivity calculations presented in the following sections. 

\section{Model-independent Limits}
\label{sec:limits}

In this section, we present the projected sensitivities of TRIDENT to DM annihilation into SM particles. The analysis follows a maximum-likelihood approach, applied separately to track and cascade neutrino event channels. 

We first describe the likelihood construction and statistical treatment, then show the resulting model-independent limits under standard assumptions for the atmospheric and astrophysical neutrino backgrounds. Finally, we assess the impact of an additional Galactic neutrino component from cosmic-ray interactions, and discuss systematic uncertainties related to the DM halo model.

\begin{figure}[h]
  \centering
  \begin{subfigure}[t]{.48\textwidth}
    \centering
    \includegraphics[width=\linewidth]{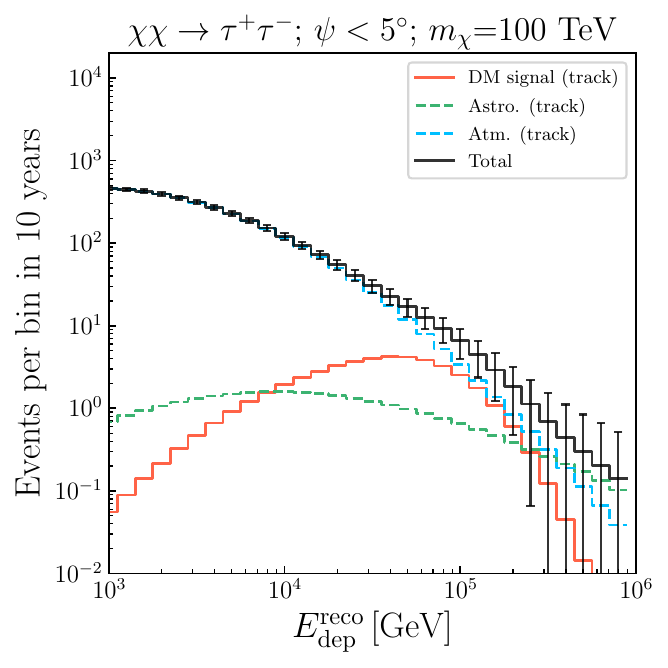}
    % \caption{Track, $\tau^+\tau^-$ ($\psi<5^\circ$)}
    \label{fig:track-tautau}
  \end{subfigure}\hfill
  \begin{subfigure}[t]{.48\textwidth}
    \centering
    \includegraphics[width=\linewidth]{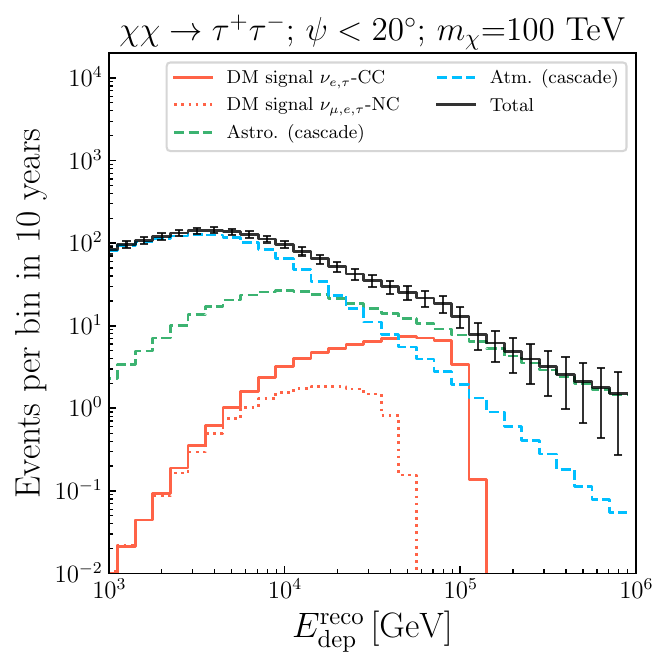}
    % \caption{Cascade, $\tau^+\tau^-$ ($\psi<20^\circ$)}
    \label{fig:cascade-tautau}
  \end{subfigure}
  
  % \vspace{0.3em} 
  
  \begin{subfigure}[t]{.48\textwidth}
    \centering
    \includegraphics[width=\linewidth]{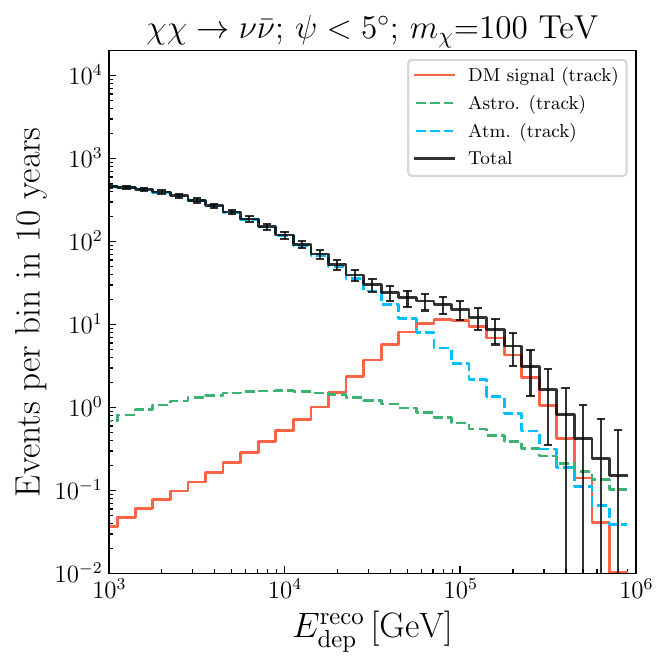}
    % \caption{Track, $\nu\bar{\nu}$ ($\psi<5^\circ$)}
    \label{fig:track-nunu}
  \end{subfigure}\hfill
  \begin{subfigure}[t]{.48\textwidth}
    \centering
    \includegraphics[width=\linewidth]{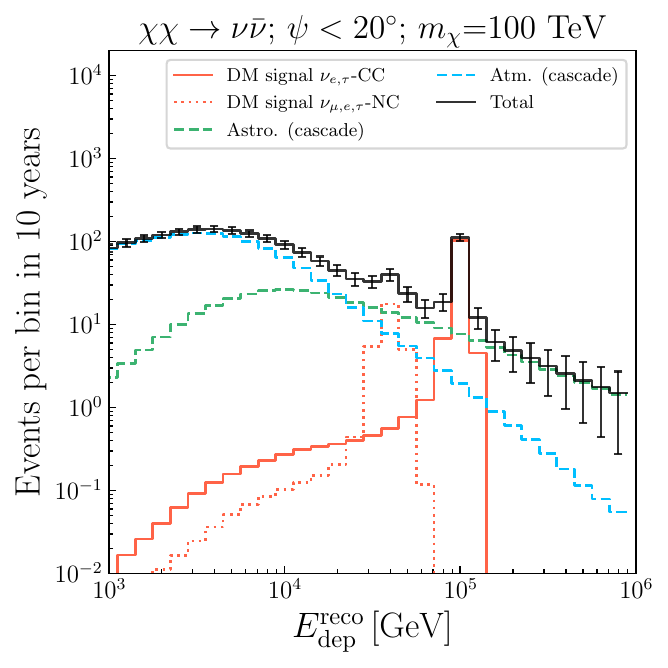}
    % \caption{Cascade, $\nu\bar{\nu}$ ($\psi<20^\circ$)}
    \label{fig:cascade-nunu}
  \end{subfigure}

  \caption{Expected reconstructed energy spectra in TRIDENT from DM annihilation channels $\chi\chi \rightarrow \tau^+\tau^-$ (top) and $\chi\chi \rightarrow \nu\bar\nu$ (bottom) for a DM mass of $m_\chi=100\,\mathrm{TeV}$. The left and right panels correspond to track events with opening angle $\psi < 5^\circ$ and cascade events with $\psi < 20^\circ$, respectively. The annihilation cross-section $\langle\sigma v\rangle$ is adopted from the current KM3NeT limits~\cite{KM3NeT:2024xca}. Distributions account for the detector’s energy reconstruction uncertainty.}
  
  \label{fig:evt_1d}
\end{figure}

\begin{figure}[h]
  \centering
  \begin{subfigure}[t]{.48\textwidth}
    \centering
    \includegraphics[width=\linewidth]{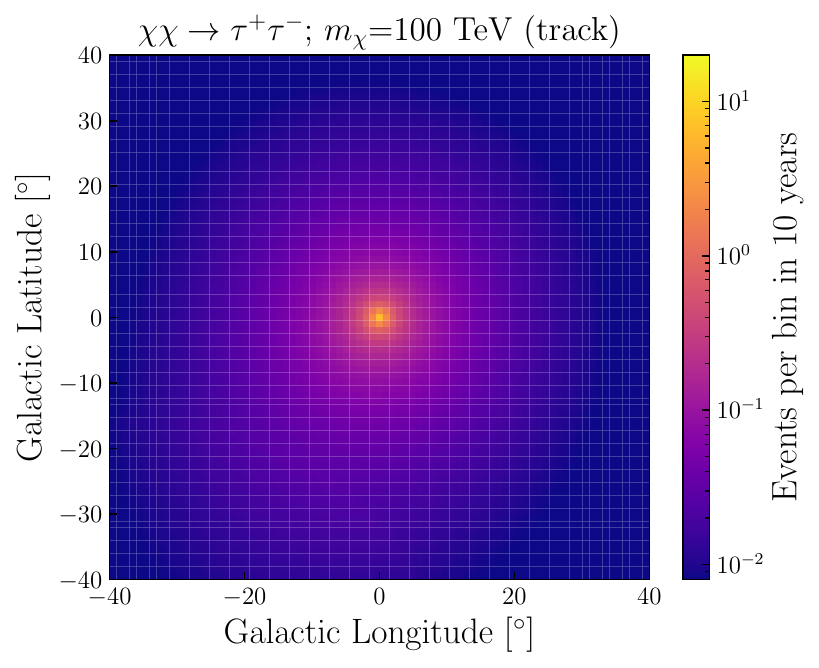}
    % \caption{Track, $\tau^+\tau^-$ ($\psi<5^\circ$)}
    \label{fig:track2d-tautau}
  \end{subfigure}\hfill
  \begin{subfigure}[t]{.48\textwidth}
    \centering
    \includegraphics[width=\linewidth]{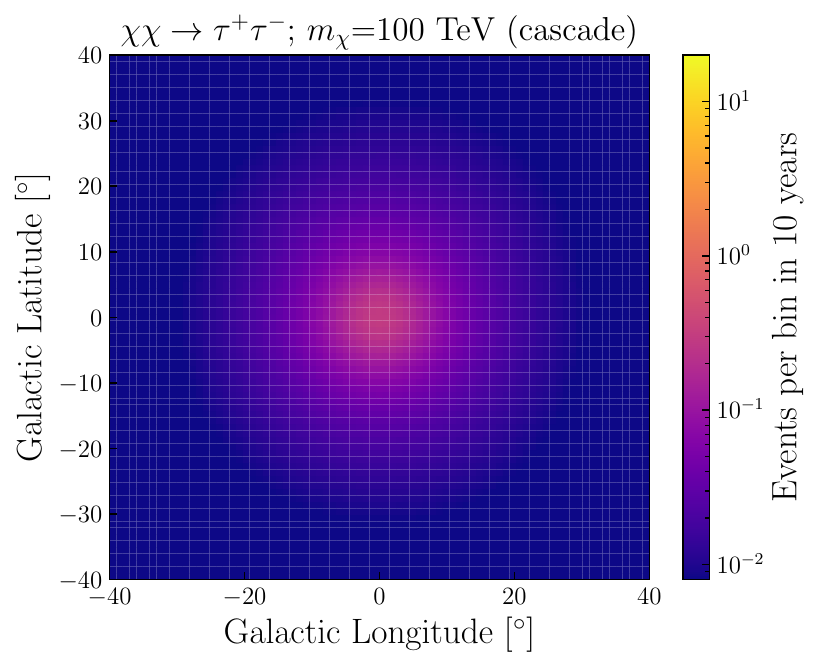}
    % \caption{Cascade, $\tau^+\tau^-$ ($\psi<20^\circ$)}
    \label{fig:cascade2d-tautau}
  \end{subfigure}
  
  % \vspace{0.3em} 
  
  \begin{subfigure}[t]{.48\textwidth}
    \centering
    \includegraphics[width=\linewidth]{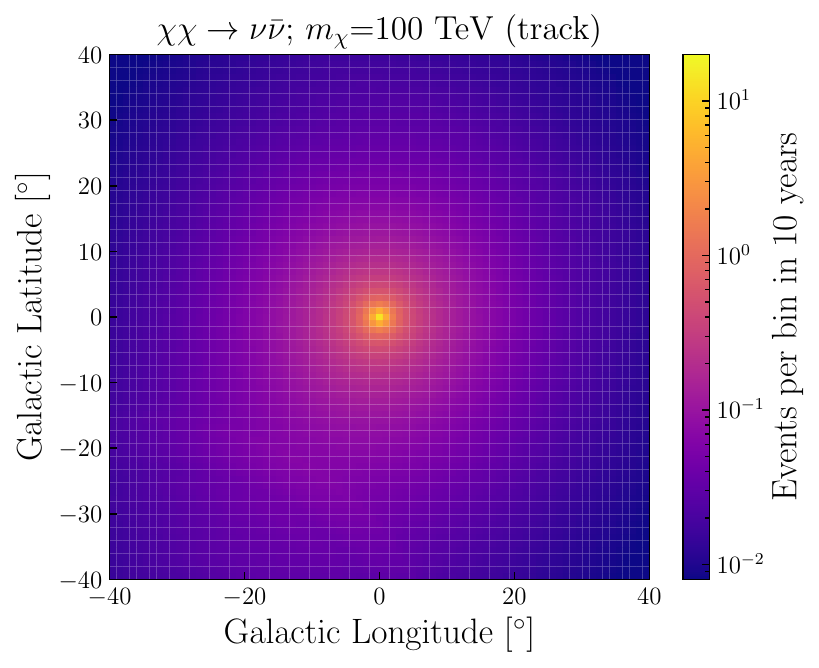}
    % \caption{Track, $\nu\bar{\nu}$ ($\psi<5^\circ$)}
    \label{fig:track2d-nunu}
  \end{subfigure}\hfill
  \begin{subfigure}[t]{.48\textwidth}
    \centering
    \includegraphics[width=\linewidth]{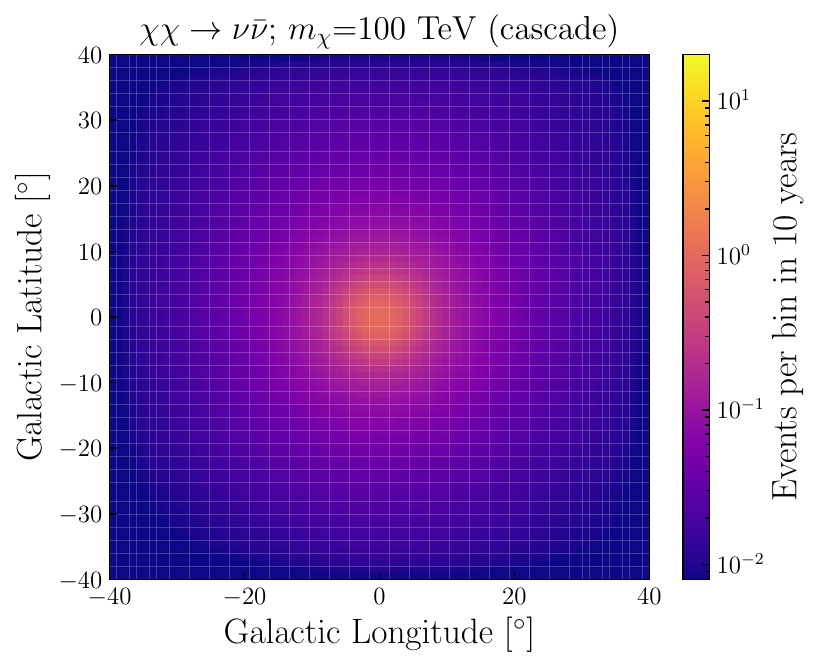}
    % \caption{Cascade, $\nu\bar{\nu}$ ($\psi<20^\circ$)}
    \label{fig:cascade2d-nunu}
  \end{subfigure}

  \caption{Reconstructed sky maps of expected DM signals in TRIDENT in Galactic coordinates, extending $\pm40^\circ$ in both latitude and longitude. The signals arise from DM annihilation channels $\chi\chi \rightarrow \tau^+\tau^-$ and $\chi\chi \rightarrow \nu\bar{\nu}$ for a DM mass of $m_\chi = 100\,\mathrm{TeV}$. The annihilation cross-section $\langle\sigma v\rangle$ is taken from the current KM3NeT limits~\cite{KM3NeT:2024xca}. The distributions account for the detector’s angular resolution.} 
  \label{fig:two_simple}
  \label{fig:evt_2d}
\end{figure}

\subsection{Likelihood Construction and Background Modelling}

We evaluate TRIDENT's sensitivity to DM annihilation signals using the track and cascade channels separately. DM signal events are extracted through the reconstructed sky map distribution of neutrino events in the Galactic Centre, where the angular resolution performance for both track and cascade events can be seen in Ref.~\cite{MortonBlake:2026allflavour}. Following the same detector-level treatment, signal and background event rates are determined using TRIDENT's calculated effective area, which implements a detector-level trigger based on a minimum number of hDOMs with coincident photons detected. Basic analysis cuts are similarly applied. The TRIDENT site latitude is taken as $17.4^\circ\mathrm{N}$. For track events only reconstructed \textit{up-going} events ($\theta_z > 90^\circ$) are considered in order to suppress the large background from atmospheric muons. The Galactic Centre region is therefore geometrically visible for approximately 60\% of the time. Cascade events conversely are selected from \textit{all directions}, as the diffuse nature of Galactic sources reduces dependence on strict angular cuts.

To account for energy reconstruction resolutions, a Gaussian smearing approximation is assumed based on preliminary energy reconstruction methods developed for the TRIDENT detector~\cite{mocen2023gnn,gnn2025prd}, applying:
$\Delta\log_{10}(E_\mathrm{reco}/E_\nu)=0.25$ for tracks and $0.04$ for cascades, corresponding to energy resolution of $\sim80\%$ and $\sim10\%$, respectively.

For an effective exposure time $T$, the expected number of events $\mu$ is computed as a convolution of the neutrino intensity $d\Phi_\nu/dE_\nu$, with the declination-dependent effective area $A_{\mathrm{eff}}(E_\nu,\delta)$, and resolution functions that describe the angular and energy reconstruction uncertainties:

\begin{equation}
    % \mu = T\int d\Omega\int dE_\nu \frac{d\Phi_\nu}{dE_\nu}A_{\mathrm{eff}}(E_\nu,\delta),
    \mu = T\int d\Omega_\nu\int dE_\nu \frac{d^2\Phi_\nu}{dE_\nu d\Omega_\nu}A_{\mathrm{eff}}\int_{\Delta E_{\mathrm{rec}}} dE_\mathrm{rec}\int_{\Delta\Omega_{\mathrm{rec}}}d\Omega_{\mathrm{rec}}\mathcal{G}_E(E_\mathrm{rec}|E_\nu)\mathcal{G}_\Omega(\Omega_{\mathrm{rec}}|E_\nu)
\end{equation}

The total expected event rate includes both \textbf{signal} and \textbf{background} components. The dominant backgrounds are:
\begin{itemize}
    \item \textbf{Atmospheric neutrinos}, modelled using the \texttt{DaemonFlux} package~\cite{daemonflux}, including both conventional and prompt components.
    \item \textbf{Atmospheric muons}, are assumed to be effectively removed through an up-going reconstructed directional cut, applied to track events only.
    \item \textbf{Diffuse astrophysical neutrinos}, modelled as an isotropic flux following IceCube’s best-fit power law~\cite{icecube2022diffuse}.
    \item \textbf{Galactic plane neutrinos}, measured by the IceCube experiment~\cite{icecubeneutrino}, are expected to be a measurable background for potential DM signal neutrinos produced in the same region. Further discussions on this background and its expected impact are found in Section~\ref{subsec:gp_bkg}.
\end{itemize}

The estimation of TRIDENT sensitivities on the upper limits of the thermally averaged DM annihilation cross-section $\langle \sigma v \rangle$ is performed using the maximum-likelihood method. The likelihood function $\mathcal{L}$ is defined as

\begin{equation}
\mathcal{L}(\langle \sigma v \rangle) = \prod_{ij} \frac{(s_{ij}+b_{ij})^{n_{ij}}}{n_{ij}!}e^{-(s_{ij}+b_{ij})},
\end{equation}

where $s_{ij}$ and $b_{ij}$ are the expected signal and background events in the $i$th energy bin and $j$th angular bin, and $n_{ij}$ is the observed count (background-only in sensitivity studies).

The expected DM signal and background for a 10-year exposure are shown in Fig.~\ref{fig:evt_1d} and Fig.~\ref{fig:evt_2d}. The event rate of the DM signal is decomposed into individual interaction components in Fig.~\ref{fig:evt_1d} as a function of reconstructed deposited energy, $E^{\rm reco}_{\rm dep}$. We take maximum opening angles of $\psi=5^\circ$ for track events and $\psi=20^\circ$ for cascade events. Since NC events yield fewer Cherenkov photons as compared to CC events, we see two peaks in the cascade energy spectrum.
Fig.~\ref{fig:evt_2d} displays the corresponding sky map of the DM signals obtained with $\psi=40^\circ$. The neutrino spectra are calculated for DM annihilation channels $\chi\chi \rightarrow \tau^+\tau^-$ and $\chi\chi \rightarrow \nu\bar{\nu}$ with a DM mass of 100~$\mathrm{TeV}$, adopting the annihilation cross-section $\langle\sigma v\rangle$ constrained by current KM3NeT limits~\cite{KM3NeT:2024xca}. 

The sensitivity for each annihilation channel hypothesis is evaluated through the fits of Asimov datasets using the test statistic, which is defined by the ratio of the maximised likelihood and the null hypothesis likelihood \cite{wilks1938large}
\begin{equation}
    \mathrm{TS} = -2 \ln \left(\frac{\mathcal{L}(\hat{s})}{\mathcal{L}(s=0)}\right).
\end{equation}
The value of cross-section $\langle \sigma v\rangle$ that yields a number of signal events $\hat{s}$ corresponding to the 95\% confidence level criterion is taken as the upper limit.

\subsection{Standard Sensitivity}
We forecast the sensitivity of TRIDENT to DM annihilation into specific SM final states, assuming a 100\% branching ratio for each channel.

\begin{figure}[h!]
  \centering
  \includegraphics[width=.48\textwidth]{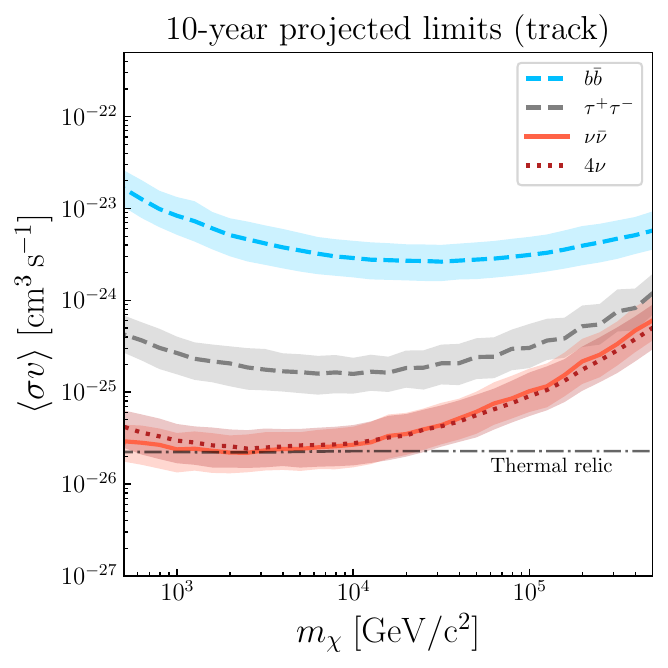}\hfill
  \includegraphics[width=.48\textwidth]{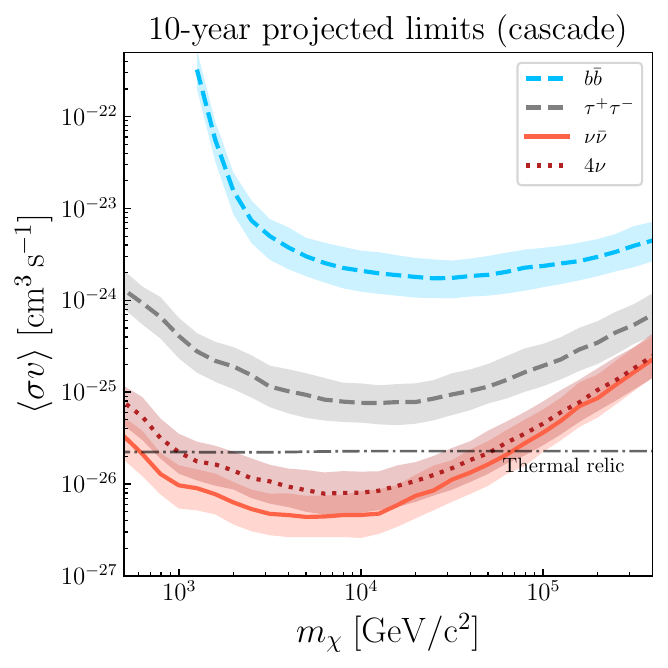}
  \caption{95\% CL upper limits of TRIDENT on the thermally averaged DM annihilation cross-section $\langle \sigma v\rangle$, as a function of the DM mass for four annihilation channels, assuming 10 years of full detector exposure}. The left (right) panel represents the sensitivity obtained from track (cascade) events.
  \label{fig:model_independent_limit}
\end{figure}

In Fig.~\ref{fig:model_independent_limit} we show the 95\% CL upper limits for the $b\bar b$, $\tau^+\tau^-$, $\nu\bar \nu$ and 4$\nu$ channels obtained from the maximum-likelihood analysis. The lines represent the median expected limits, while the colour bands denote the 68\% intervals around the expected value, taking into account the statistical uncertainties from Monte Carlo simulations. The sensitivity derived from track events is shown in the left panel, whereas the right panel depicts the sensitivity obtained from cascade events.

Overall, the limits exhibit a clear hierarchy among the annihilation channels. The $b\bar b$ and $\tau^+\tau^-$ final states yield comparatively weaker sensitivities, primarily because these channels produce broader secondary particles and softer neutrino spectra. In particular, for the $b\bar{b}$ channel, the likelihood fits based on cascade events fail to converge at dark matter masses below 1 $\mathrm{TeV}$, which is attributed to the limited effective area for cascade events at low energies. The neutrino final-state channels ($\nu\bar\nu$ and 4$\nu$) provide the strongest sensitivities as expected, since annihilation directly into neutrinos produces a sharply peaked and hard neutrino spectrum. Both analyses based on tracks and cascades are expected to reach the thermal relic value around multi-$\mathrm{TeV}$ for $\nu\bar\nu$ and $4\nu$ channels. The sensitivities of cascades benefit from the better energy resolution, which provides a boost for all the channels. The relative behaviour of the 4$\nu$ and $\nu\bar\nu$ channels diverges between the two event topologies. For tracks, the limits for 4$\nu$ nearly overlap with those of $\nu\bar\nu$, since track events have relatively poor energy resolution and are thus less sensitive to the differences in the injected neutrino spectrum, leading to comparable sensitivities. In contrast, the more precise energy reconstruction of cascades makes the spectra from box-shaped 4$\nu$ flux and line-like $\nu\bar\nu$ flux distinguishable. As a result, the cascade channel achieves better sensitivities for the $\nu\bar\nu$ channel against the 4$\nu$ channel. To assess the impact of electroweak corrections, we also calculated the sensitivities for the neutrino final states using the corresponding naive spectra. We found that the effect of electroweak radiation is negligible for $m_\chi \lesssim 2\,{\rm TeV}$. Even in the high-mass regime $m_\chi \gtrsim 100\,{\rm TeV}$, the degradation of the sensitivity remains less than a factor of two.

\subsection{Galactic Plane Neutrino Background}
\label{subsec:gp_bkg}

In addition to the previously considered atmospheric and diffuse astrophysical neutrinos, Galactic neutrino emission originating from the interactions of hadronic cosmic rays with interstellar gas is also expected \cite{Gaisser:1994yf, IceCube:2014rwe, icecubeneutrino}. This background is not very well understood, and introduces an additional uncertainty. To reflect possible novel features at high energy, we follow the modelling of Refs.~\cite{Kaci_2025, microquasars, neutrinos_with_trident}, where cosmic rays are injected by individual sources with a discrete distribution of positions. This approach is more physically motivated than the use of propagation codes, such as GALPROP \cite{galprop} and DRAGON \cite{dragon1, dragon2}, which assume a smooth distribution of cosmic-ray sources in the Galaxy. The shorter confinement time of $\rm{TeV}-\rm{PeV}$ cosmic rays, compared to $\rm{GeV}$ counterparts, leads to more visible imprints from individual sources. TRIDENT sensitivities to DM should also be re-evaluated without prior knowledge of the expected distribution of the Galactic neutrinos.

Galactic neutrinos contribute in the form of two components, diffuse Galactic neutrino emission and Galactic neutrino source emission. The former arises from hadronic interactions of the large-scale Galactic cosmic-ray population with interstellar gas. This is expected to be concentrated along the Galactic plane and roughly following the gas distribution up to energies of a few tens of $\rm{TeV}$ (taken from Refs.~\cite {Mertsch:2020qld, Mertsch:2022oee} for this work), while for higher energies, this smooth approximation becomes less accurate because the short residence time of PeV cosmic rays makes the neutrino emissivity sensitive to the stochastic distribution of recent Galactic accelerators~\cite{Kaci_2025}. This is when Galactic neutrino source emission can be important. To model this we use Monte-Carlo realisations of discrete PeVatron-like supernova remnants with a total energy budget of $10^{51}\,{\rm erg}$, 10\% of which goes into cosmic-ray acceleration. Sources are generated with random ages at a rate of 3 per century, with Galactocentric radii sampled from the axisymmetric surface density \cite{snr_dist2013} and vertical positions uniformly distributed within the Galactic disk. The number of PeVatron sources in each realisation is determined by the birth rate, the sampled age range ($10\,\rm{Myr}$ in this work), and the assumed fraction of generated remnants that accelerate cosmic rays up to PeV energies. The injected cosmic-ray spectrum is assumed to follow a power law with index $2.4$ which becomes $\simeq2.7$ after propagation. The cosmic rays are propagated using Green's function solutions of the diffusion equation, and the resulting neutrino emission is obtained by convolving the propagated cosmic-ray density with the interstellar gas distribution.
The overall normalisation of the neutrino flux is adjusted to reproduce the IceCube measurement around $70 \, \rm{TeV}$ averaged over the sky using the $\pi^0$ template \cite{icecubeneutrino}, note that IceCube are working on an updated analysis~\cite{Neste:2025HU}.

\begin{figure}[h]
  \centering
  \includegraphics[width=.48\textwidth]{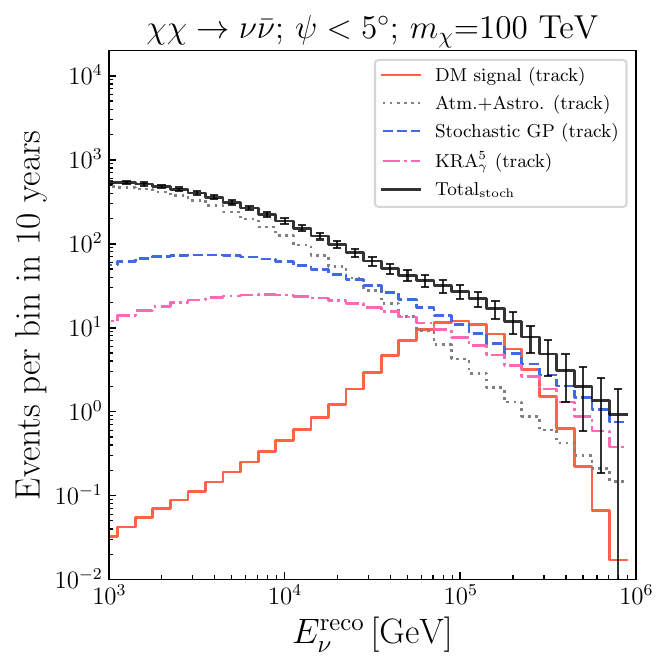}\hfill
  \includegraphics[width=.48\textwidth]{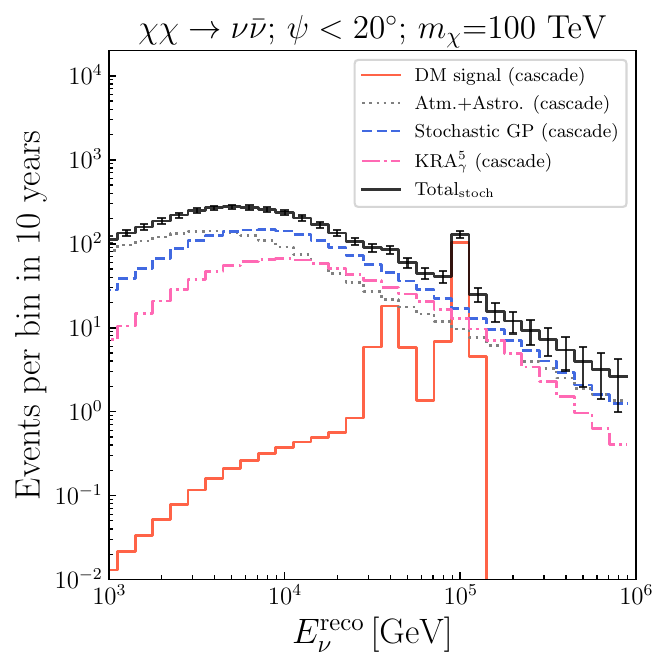}
  \caption{The event rate of Galactic plane neutrinos background from one stochastic Galactic plane template and KRA$_\gamma^5$ model, together with the DM signal from $\chi\chi\rightarrow \nu\bar\nu$ channel and the combined atmospheric and diffuse astrophysical background, for both track (left) and cascade (right) events. The Galactic contribution included in the total events curve is the stochastic Galactic-plane neutrino component.}
  \label{fig:eventgp}
\end{figure}

Our analysis method has to account for the stochastic nature of this background, we adopt the following procedure. First, we randomly generate different realisations of the Milky Way, where the Galactic neutrino emission is generated by different sets of Galactic sources drawn randomly for each simulation. This yields sky maps of the Galactic neutrino flux with stochastic spatial distributions. In the sensitivity calculation, some maps are randomly selected to form Asimov datasets for the potential DM signal, while others are used as probability distribution functions. We adopted 20 sky-map realisations and the average sensitivity is taken. For comparison, we also include the KRA$_\gamma^5$~\cite{kra52015} model, a gamma-ray-constrained Galactic diffuse-emission template with a 5 PeV cosmic-ray cutoff. The template is normalised using the best-fit flux in the IceCube observation of high-energy neutrino emission from the Galactic plane~\cite{icecubeneutrino}.  We chose to use KRA$_\gamma^5$ as opposed to other backgrounds such as $\pi^0$~\cite{kra52015} because it is more peaked around the Galactic Centre and will therefore have a bigger effect. Unlike the baseline stochastic model, which is based on discrete realisations of Galactic cosmic-ray sources, KRA$_\gamma^5$ provides a more spatially continuous neutrino emission from the Galactic Centre region. Fig.~\ref{fig:eventgp} shows the Galactic neutrino event rate obtained from one stochastic Galactic plane sky map and from the KRA$_\gamma^5$ model, together with the DM signal from $\chi\chi\rightarrow \nu\bar\nu$ channel and the atmospheric combined with diffuse astrophysical background contributions.

\begin{figure}[h]
    \centering
    \includegraphics[width=0.5\linewidth]{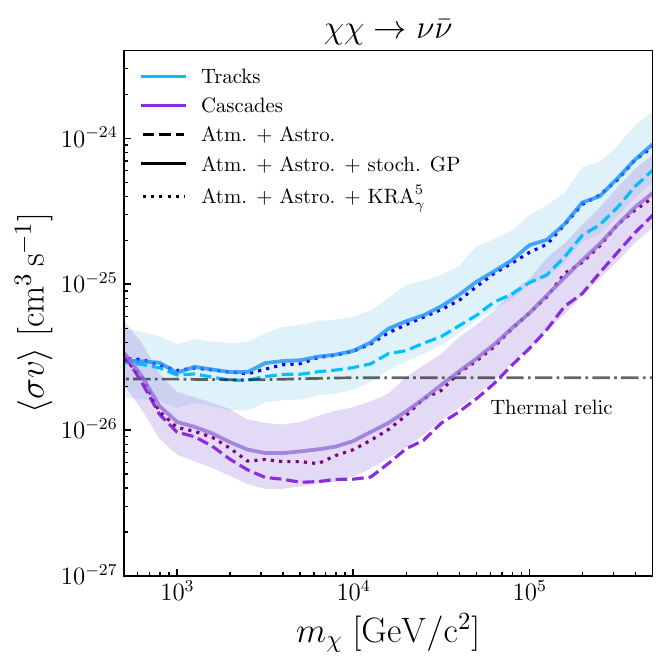}
    \caption{DM signal sensitivity in the $\chi\chi\rightarrow \nu\bar{\nu}$ channel for track (cyan) and cascade (purple) events. Results are shown without the Galactic-plane background (dashed line) and with its impact modelled using stochastic Galactic-plane maps (solid line) or the $\mathrm{KRA}^5_\gamma$ template (dotted line). We assume 10 years of full detector exposure. Coloured bands represent 68\% uncertainty intervals.}
    \label{fig:galactic_plane}
\end{figure}

The impact of the Galactic neutrinos on the sensitivity for the representative $\nu\bar \nu$ channel is shown in Fig.~\ref{fig:galactic_plane}, which presents the sensitivity to DM as a function of DM mass for track (cyan) and cascade events (purple), including and excluding Galactic neutrino emission, modelled using either the stochastic or the KRA$_\gamma^5$ template. The central 68\% interval is shown as colour bands, indicating the stochastic effect from various realisations. Across the ensemble, the resulting sensitivity varies by approximately a factor of two. We find that the Galactic-plane neutrino background degrades TRIDENT's sensitivity in both event channels above $\sim 1 \,{\rm TeV}$, with the strongest impact occurring above $\sim 10\,{\rm TeV}$, where the hard Galactic-plane spectra become dominant. Moreover, the impact is more pronounced for cascade events, as their poorer angular resolution makes it more difficult to separate the Galactic-plane background from the DM signal toward the Galactic Centre region. We have also evaluated the $b\bar b$, $\tau^+\tau^-$ and $4\nu$ annihilation channels, which we show in Appendix~\ref{app:GP}. The corresponding impact of the Galactic-plane neutrino background follows a similar qualitative behaviour to that observed for the $\nu\bar\nu$ channel.

\subsection{Dark Matter Halo Model }
\label{subsec:dmhalo}

In addition to the neutrino backgrounds mentioned above, substantial uncertainties of the DM density profile can significantly limit TRIDENT's sensitivity to $\langle\sigma v\rangle$. Efforts to understand the potential density profile primarily come from astrometry and N-body cosmological simulations. 
Modern N-body simulations such as Auriga L3~\cite{Grand:2024xnm}, FIRE-2~\cite{Wetzel:2022man}, VINTERGATAN-GM~\cite{Rey:2022mwh}, TNG50~\cite{Pillepich:2019bmb}, incorporate baryonic effects. Furthermore, in the era of survey telescopes such as Gaia \cite{vallenari2023gaia,GaiaDR1,GaiaDR2,GaiaEDR3}, astrometry studies are also able to provide competitive constraints on the profile in the Milky Way~\cite{Zhou:2022lar, ou2024dark}. 

\begin{figure}[h]
  \centering
  \includegraphics[width=.48\textwidth]{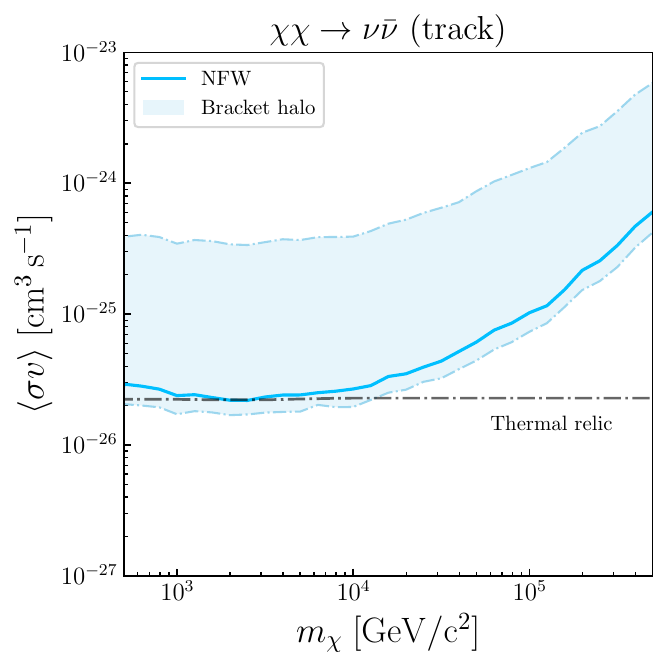}\hfill
  \includegraphics[width=.48\textwidth]{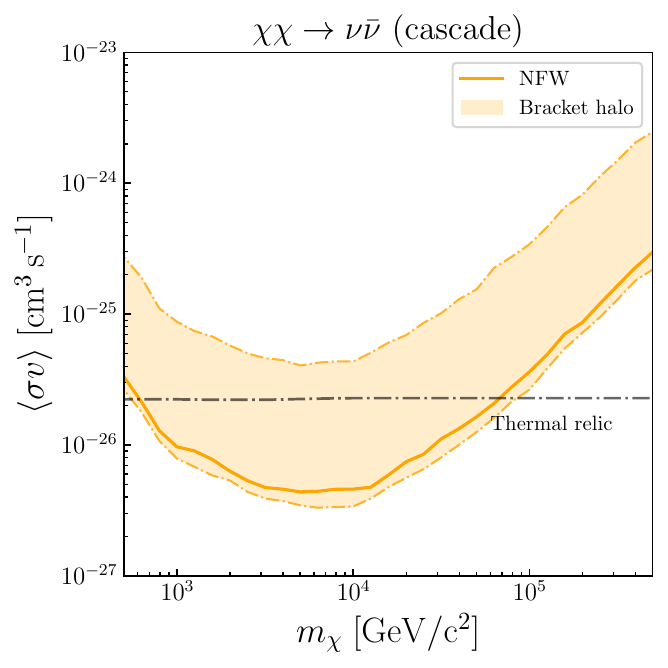}
  \caption{The expected sensitivity of the $\chi\chi\rightarrow \nu\bar\nu$ channel for both track (left) and cascade (right) events as in Fig.~\ref{fig:model_independent_limit} but showing the effects from variations in the halo model. The solid curves indicate the limits using the NFW profile as a benchmark, while the coloured bands represent the variations from the Bracket halo model.}
  \label{fig:bracket}
\end{figure}

Despite substantial progress, the uncertainty in the profile remains large, especially in the central regions. Ref.~\cite{Hussein:2025xwm} proposed using a `bracketed' profile, which takes the state-of-the-art results from N-body simulations and astrometry in order to quantify our ignorance, see Fig.~2 of Ref.~\cite{Hussein:2025xwm}. The profile maintains an uncertainty slightly below one order of magnitude in the inner region ($\lesssim 1\,{\rm kpc}$), but it is still sufficient to disfavour profiles with a large core. In Fig.~\ref{fig:bracket} we show the effect this bracketed halo function has on our projected $\nu\bar\nu$ channel sensitivity; panels left and right show our results for track and cascade signals, respectively. We see that for both tracks and cascades, the TRIDENT sensitivity is affected by around an order of magnitude for all dark matter masses. We performed the same analysis for the $b\bar b$, $\tau^+\tau^-$ and $4\nu$ annihilation channels and find qualitatively similar behaviour: cuspy halo profiles improve the sensitivity to $\langle\sigma v\rangle$ relative to the NFW profile by a small percentage, whereas cored profiles degrade the sensitivity by a factor of 10 for tracks and by a factor of 10–20 for cascades. This factor is fairly stable over the entire DM mass range. Comparing Fig.~\ref{fig:model_independent_limit} with Fig~\ref{fig:galactic_plane} and Fig.~\ref{fig:bracket}, we can see that the halo uncertainty remains the leading cause of uncertainty in TRIDENT's sensitivity to DM annihilations.

\section{Particle Model Inference and Complementarity}
\label{sec:particle_model}

\begin{figure}[h!]
    \centering
    \includegraphics[width=0.7\linewidth]{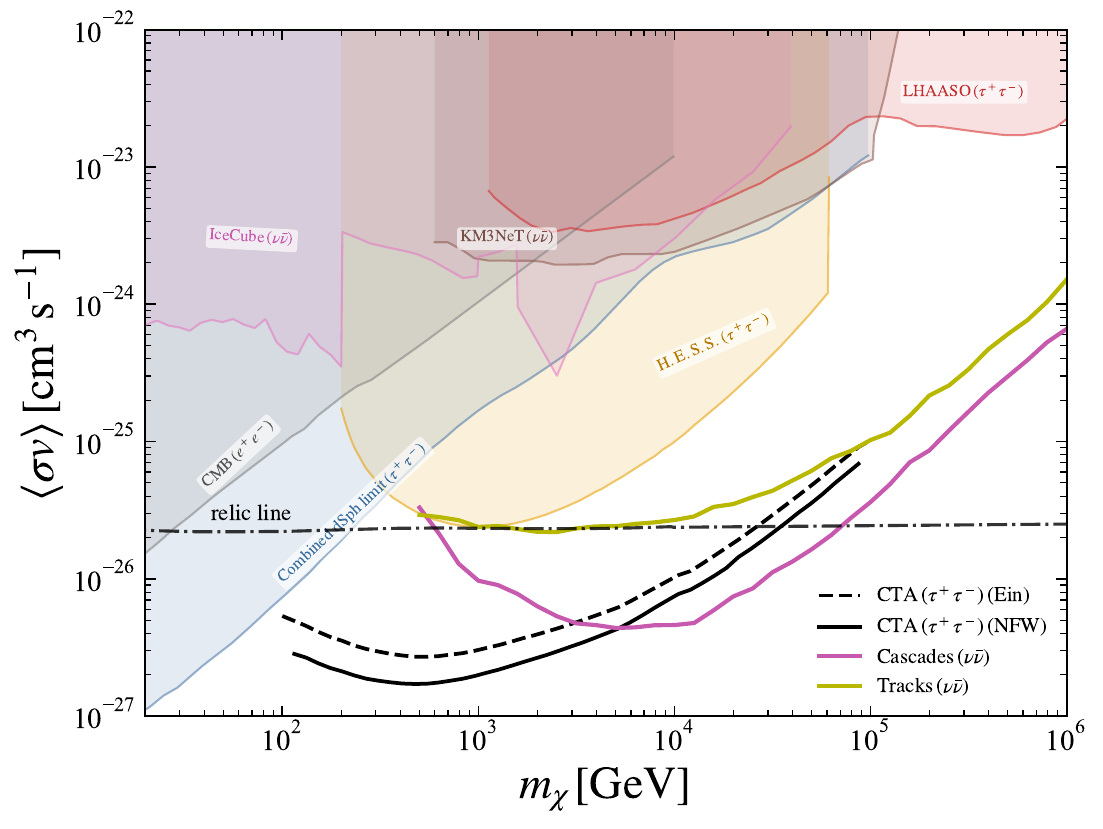}

    \caption{
    Landscape of current constraints and projected sensitivities on the thermally averaged DM annihilation cross-section $\langle \sigma v\rangle$ as a function of the DM mass $m_\chi$.
    The coloured shaded regions show existing upper limits from indirect searches, including the combined dwarf spheroidal galaxy limits~\cite{Fermi-LAT:2025gei} from Fermi-LAT, HAWC, H.E.S.S., MAGIC and VERITAS, the Galactic Centre limits from H.E.S.S.~\cite{HESS:2022ygk}, the diffuse gamma-ray limits from LHAASO~\cite{LHAASO:2024upb}, the limits from IceCube~\cite{IceCube:2023ies,IceCube:2025fcn} and KM3NeT~\cite{KM3NeT:2024xca}, and the CMB constraint from Planck~\cite{Slatyer:2015jla}. For each experiment we chose the most constraining channel which we signify in the bracket next to the label. 
    The black dash-dotted curve denotes the annihilation cross-section required to reproduce the observed DM relic abundance through standard thermal freeze-out~\cite{Steigman:2012nb}.
    The projected CTA sensitivity to the $\tau^+\tau^-$ channel is shown for both Einasto and NFW halo profiles, indicated by the dashed and solid black curves, respectively.
    The magenta and olive solid curves show the projected TRIDENT sensitivities for cascade-like and track-like events, respectively, assuming a 100\% annihilation branching fraction into neutrinos.
    }
    \label{fig:sigv_context}
\end{figure}

In the previous sections, we demonstrated that the TRIDENT experiment can probe the DM thermal relic benchmark parameters under the assumption of a 100\% annihilation rate into neutrinos. However, this is not a particularly likely scenario because neutrinos reside in left-handed SU(2) doublets, meaning many DM models with annihilations into neutrinos also have comparable annihilations into charged leptons and SM gauge bosons. To better contextualise our projections in the landscape of indirect detection searches, we show, in Fig.~\ref{fig:sigv_context} a non-exhaustive plot of the current limits on the dark matter annihilation cross-section. We show the most constraining channels for all analyses. In shaded blue we show the combined dwarf spheroidal analysis performed by Fermi-LAT~\cite{Fermi-LAT:2015att}, HAWC~\cite{HAWC:2017mfa}, H.E.S.S.~\cite{HESS:2018kom}, MAGIC~\cite{MAGIC:2021mog} and VERITAS~\cite{VERITAS:2017tif} presented in Ref.~\cite{Fermi-LAT:2025gei}. We show the most constraining $\tau^+\tau^-$ channel, taking the more constraining J-factor set~\cite{Bonnivard:2014kza,Bonnivard:2015xpq}. In shaded red we show the dwarf spheroidal result from LHAASO~\cite{LHAASO:2024upb} ($\tau^+\tau^-$). The H.E.S.S. constraint is their Galactic Centre constraint, also for $\tau^+\tau^-$~\cite{HESS:2022ygk}. Since the H.E.S.S. results for the relevant annihilation channels assume an Einasto profile, we estimate the H.E.S.S. NFW limit by rescaling according to the relative $W^+W^-$ sensitivities for Einasto and NFW profiles. This weakens the limit somewhat but it makes the comparison between our result more meaningful. We make a similar approximation for the CTA projections~\cite{CTA:2020qlo}, however we see the NFW projections are stronger. The PLANCK limit is strongest for $e^+e^-$~\cite{Slatyer:2015jla}. We show the existing limits on the $\nu\bar{\nu}$ Galactic Centre for IceCube~\cite{IceCube:2025fcn} and KM3NeT~\cite{KM3NeT:2024xca}. 

In Fig.~\ref{fig:sigv_context} we can see TRIDENT will exhibit excellent complementarity with future $\gamma$-ray searches for dark matter. In particular we see that the cascade signal is vitally important for this. Furthermore, the impact of TRIDENT's sensitivity depends on whether there exist well-motivated particle physics models that are not already excluded. 

Although many models are available~\cite{ElAisati:2017ppn, Blennow:2019fhy,Kopp:2014tsa,Ibarra:2015fqa,Arina:2020udz, Arina:2025zpi}, we take as an example the U(1)$_{L_i-L_j}$ extension to the SM~\cite{He:1991qd,He:1990pn}. This model is theoretically consistent, simple, and can accommodate DM. It was shown in Ref.~\cite{BasegmezDuPree:2021fpo} that high-energy neutrino telescopes would be able to provide particularly competitive limits~\cite{Bernal:2025szh}. This is primarily because the annihilation rate to neutrinos is relatively high, even when $m_{\chi}>1\,{\rm TeV}$, an energy region where $\gamma$-ray telescopes lose sensitivity. Furthermore, the model does not couple to quarks at tree-level, weakening bounds from DM direct detection and colliders. Owing to TRIDENT’s large exposure to the Galactic Centre and its excellent angular resolution, we expect it to provide leading sensitivity in this class of models. In this section we will show that TRIDENT is expected to probe regions of parameter space that are currently unconstrained, including the parameter regions that correspond to thermal freeze-out DM. 

Before we describe the specifics of the particle model, we briefly review how we calculate the relevant DM properties used later. First, when calculating the thermally averaged cross-section during thermal freeze-out, we use~\cite{Gondolo:1990dk,Bauer:2017qwy}
\begin{equation}
    \langle \sigma v \rangle =\frac{2\pi^2 T}{\left(4\pi m_\chi^2 T K_2(m_\chi/T)\right)^2} \int_{4m_\chi^2}^\infty ds \sqrt{s} (s - 4m_\chi^2) K_1 \left( \frac{\sqrt{s}}{T} \right) \sigma_{\chi \chi \to f f}(s)\,,\label{eq:cosmic_sigv}
\end{equation}
assuming the Maxwell-Boltzmann distribution of the DM phase space. Here, $s$ is the Mandelstam variable, $K_1$ and $K_2$ are the modified Bessel functions of the second kind, $T$ is the temperature, which at freeze-out, we determine according to Ref.~\cite{Steigman:2012nb}. The cross-section $\sigma_{\chi\chi\to ff }$ is the total annihilation cross-section. 

To calculate $\langle\sigma v\rangle$ in the astrophysical environments, one replaces the phase space distributions in Eq.~(\ref{eq:cosmic_sigv}) with the DM velocity distribution. Properly normalised, one can perform the following rearrangement~\cite{Ferrer:2013cla,Ambrogi:2018jqj}
\begin{equation}
\langle \sigma v \rangle_i = \int \mathrm{d}v_{\mathrm{rel}} \,
\tilde{f}_{\mathrm{rel}}(v_{\mathrm{rel}}) \, \sigma_iv_{\mathrm{rel}} \, ,
\label{eq:gal_sigv}
\end{equation}
where $v_{\rm rel}$ is the relative DM velocity, and the subscript of $\sigma$ signifies a specific annihilation channel. The velocity distribution function can be simplified if one assumes the Maxwell-Boltzmann distribution,  
\begin{equation}
\tilde{f}_{\mathrm{rel}}(v_{\mathrm{rel}}) \equiv
4\pi v_{\mathrm{rel}}^{2} \int \mathrm{d}^{3}v_{\mathrm{CM}} \,
f(\mathbf{v}_{\mathrm{CM}} + \mathbf{v}_{\mathrm{rel}}/2) \,
f(\mathbf{v}_{\mathrm{CM}} - \mathbf{v}_{\mathrm{rel}}/2) \, =
\sqrt{\frac{2}{\pi}} \, \frac{v_{\mathrm{rel}}^{2}}{v_0^{3}}
\exp\!\left(-\frac{v_{\mathrm{rel}}^{2}}{2v_0^{2}}\right) \, ,
\end{equation}
where $v_0$ is the most probable velocity, which we take to be $220\,{\rm km}\,{\rm s}^{-1}$~\cite{Mao_2013, McMillan:2016jtx, Bozorgnia:2017brl} for the Galactic Centre and $10\,{\rm km}\,{\rm s}^{-1}$~\cite{Walker_2007,Walker:2012td} for dwarf spheroidal galaxies. To map the indirect limits for a set of given annihilation channels, $\sigma^{\rm UL}_i$,  onto our model parameter space we determine the combined upper limit using model branching ratios ${\rm Br}_i$, 
\begin{equation}
    \langle\sigma v\rangle_{\rm tot}^{\rm lim}=\left(\sum_i \frac{{\rm Br}_i}{\langle\sigma v\rangle_i^{\rm lim}}\right)^{-1}\quad {\rm where}\quad {\rm Br}_i=\frac{{\langle\sigma(\chi\bar{\chi}\to i)\,v\rangle}}{\langle\sigma_{\rm tot}\,v\rangle}\,.
\end{equation}
We will show our TRIDENT projection as a combination of the track and cascade analyses above. For $\gamma$-ray telescopes, the most constraining bounds above $\sim 200\,{\rm GeV}$ come from the H.E.S.S. Galactic Centre analysis~\cite{HESS:2022ygk}. 
%\sout{As shown in Section~\ref{subsec:dmhalo}, the Galactic Centre is subject to a large uncertainty due to the DM halo.} 
%\sout{Since the H.E.S.S. results for the relevant annihilation channels assume an Einasto profile~\cite{HESS:2022ygk}, we estimate the H.E.S.S. NFW limit by rescaling according to the relative $W^+W^-$ sensitivities assuming the Einasto and NFW profiles, respectively.} 

Finally we will also make use of the latest results from direct DM detection experiments. We will explicitly show the PandaX-4T result~\cite{PandaX:2024qfu}, but other xenon-based experiments such as LZ~\cite{LZ:2024vge} and XENONnT~\cite{XENON:2025vwd} have achieved similar levels of sensitivity.

\subsection{The $L_i-L_j$ Model}

The U(1)$_{L_i-L_j}$ extension of the SM is one of the simplest theoretically consistent extensions available. The model introduces a new U(1)$_{L_i-L_j}$ boson, $Z^\prime$, which couples to the SM via the following current
\begin{equation}
    j_{\alpha}^{L_i-L_j}=\bar{L}_i\gamma_\alpha L_i + \bar{\ell}_i\gamma_\alpha \ell_i -\bar{L}_j\gamma_\alpha L_j - \bar{\ell}_j\gamma_\alpha \ell_j,
\end{equation}
where $L$ and $\ell$ are the left-handed and right-handed lepton fields, respectively, the subscript denotes the flavour and $i\neq j$. For a DM candidate we introduce a Dirac DM particle, $\chi$, which must interact with $Z^\prime$ in a vector-like way to maintain anomaly freedom~\cite{Barr:1986hj,He:1991qd,He:1990pn,Dror:2017ehi,Ellis:2017tkh}. The dark sector current is then 
\begin{equation}
    j_\alpha^{\chi}=\bar{\chi}\gamma_\alpha\chi= \bar{\chi}_L\gamma_\alpha\chi_L +\bar{\chi}_R\gamma_\alpha\chi_R\, ,
\end{equation}
such that the interaction terms are
\begin{equation}
\mathcal{L}_{\rm int} =\;
\frac{\epsilon}{2 \,\cos{\theta_W}}\, B_{\alpha\beta} Z^{\prime\,\alpha\beta} \nonumber 
+ g_{Z^\prime}^{\rm DM} Z^{\prime\,\alpha}\,j_{\alpha}^{\chi}
+ g_{Z^\prime}^{\rm SM}\,  Z^{\prime\,\alpha}\, j_{\alpha}^{L_i-L_j} \,,
\label{eq:L_U1}
\end{equation}
where the $B_{\alpha\beta}$ is the SM hypercharge field strength tensor and $\theta_W$ is the weak mixing angle. We have also introduced couplings for the kinetic mixing, $\epsilon$~\cite{Okun:1982xi,Holdom:1985ag},  and the DM and SM currents, $g_{Z^\prime}^{\rm DM}$ and $g_{Z^\prime}^{\rm SM}$ respectively. We have normalised the kinetic mixing such that in the electroweak broken phase, $\epsilon$ can be determined by loop mixing between the photon and the new boson $Z^\prime$~\cite{Araki:2017wyg,Bauer:2022nwt}, 
\begin{equation}
    \epsilon(p\to0)=\frac{e \,g_{Z^\prime}^{\rm SM}}{6\pi^2}\ln\left(\frac{m_j}{m_i}\right) \label{eq:epsilon}
\end{equation}
where $e$ is the electric charge. Depending on the model flavour, the size of the kinetic mixing will vary slightly, for $L_e-L_\mu$, $L_e-L_\tau$, $L_\mu-L_\tau$, we have $\epsilon\sim g_{Z^\prime}^{\rm SM}/30$,  $\epsilon\sim g_{Z^\prime}^{\rm SM}/20$, and $\epsilon\sim g_{Z^\prime}^{\rm SM}/60$ respectively. In the figures below, we show the results for the $L_\mu -L_\tau$ model and comment on the differences for the other $L_i-L_j$ models. 

When $m_{Z^\prime}>m_{\chi}$, DM annihilation proceeds through s-channel scattering into charged leptons and neutrinos with a branching ratio of $\sim 2/3$ and $\sim 1/3$, respectively. 
When $m_\chi>m_{Z^\prime}$, the $\chi\bar{\chi}\to Z^\prime Z^\prime$ annihilation opens up and can be important. For a neutrino telescope this process will be detectable via the decay particles from the $Z^\prime$, leading to the box-shaped $4\nu$ spectra described in Section~\ref{sec:dmflux}, for which the projected TRIDENT sensitivities were shown in Fig.~\ref{fig:model_independent_limit}. For such channels we determine the branching ratio by 
\begin{equation}
    {\rm Br}_{i\bar{i}\,j\bar{j}}={\rm Br}_{Z^\prime Z^{\prime}} \times\left(\frac{\Gamma\left(Z^\prime\to i\bar{i}\right)}{\Gamma_{\rm tot}}\right)\times\left(\frac{\Gamma\left(Z^\prime\to j\bar{j}\right)}{\Gamma_{\rm tot}}\right)\,, 
\end{equation}
where we multiply by $2$ if $i\neq j$. For the $\chi\bar{\chi}\to Z^\prime Z^\prime\to\nu\bar{\nu}\ell^+\ell^-$ channel, we estimate the sensitivity by rescaling the $4\nu$ sensitivity by a factor of $2$ to reflect the reduced signal spectra ${\rm d}N_{\nu}/{\rm d}E_\nu$.

\begin{figure}
    \centering
    \includegraphics[width=\linewidth]{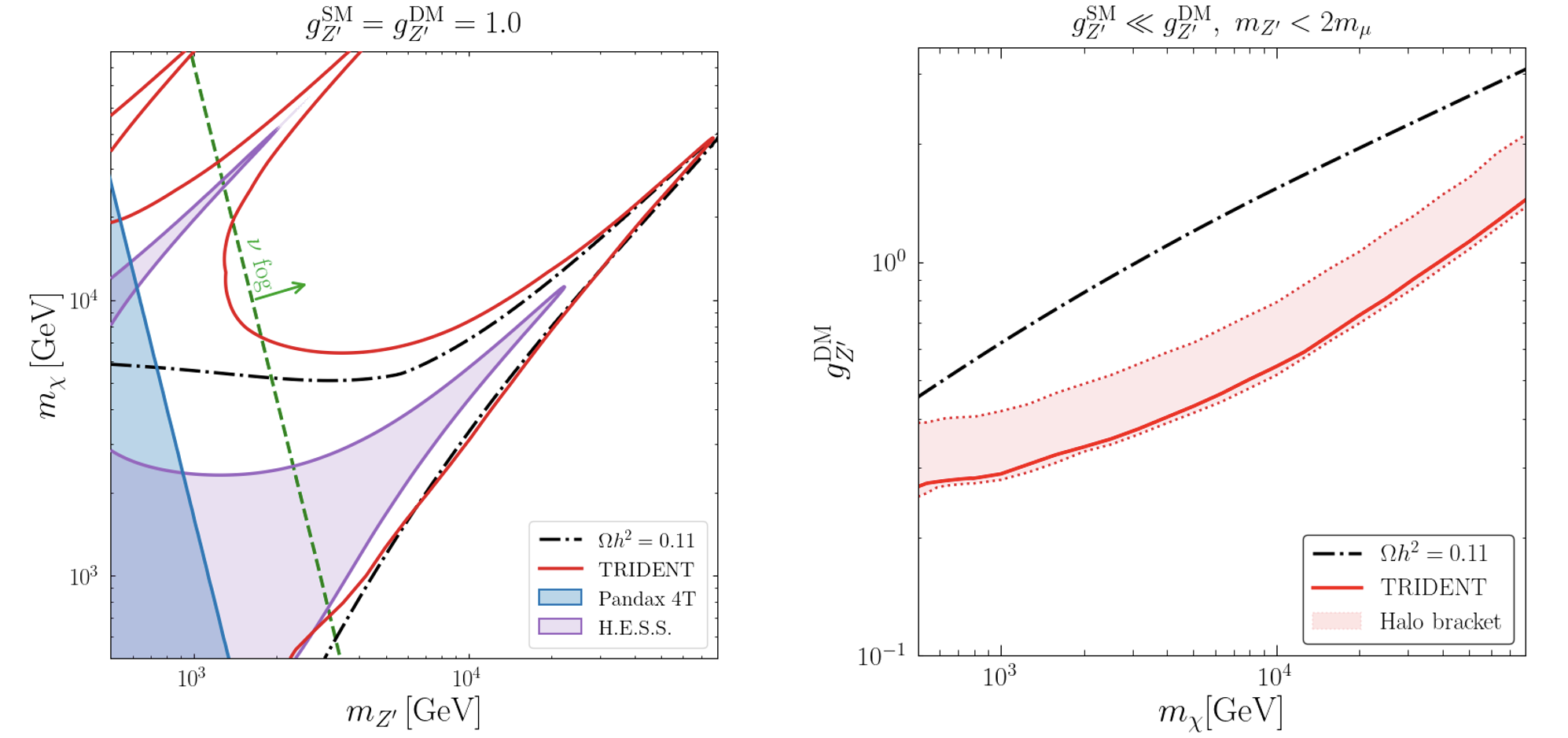}
    \caption{Plots showing the reach of TRIDENT on the model we consider here. Left panel shows the $\left(m_{Z^\prime}, m_\chi\right)$ plane assuming $g^{\rm SM}_{Z^\prime}=g^{\rm DM}_{Z^\prime}=1.0$. We show the strongest limits from $\gamma$-ray telescopes (H.E.S.S.~\cite{HESS:2022ygk}) and direct detection (PandaX-4T~\cite{PandaX:2024qfu}) with the purple and blue shaded regions, respectively. The neutrino fog~\cite{O_Hare_2021} is shown by the green dashed line. Right panel shows the TRIDENT sensitivity in the $\left(m_\chi, \,g^{\rm DM}_{Z^\prime}\right)$ plane, now assuming $g^{\rm SM}_{Z^\prime}\ll g^{\rm DM}_{Z^\prime}$. In both panels we show the calculated relic line cast from Ref.~\cite{Steigman:2012nb}.}
    \label{fig:Lmu_Ltau_model}
\end{figure}

We show our projected TRIDENT limit on the parameter space in the left panel of Fig.~\ref{fig:Lmu_Ltau_model} in red, assuming $g_{Z^\prime}^{\rm DM}=g_{Z^\prime}^{\rm SM}=1.0$ and an NFW DM profile. One can see that the TRIDENT sensitivity will be able to cover regions near or above the thermal relic line. When $m_{Z^\prime}<m_{\chi}$ the t-channel $\chi\bar{\chi}\to Z^\prime Z^\prime$ process can become important, ${\rm Br}_{Z^\prime Z^{\prime}}\sim 0.25$ of the total annihilation cross-section. In the same regions of parameter space, the relatively light $Z^\prime$ compared to $\chi$ can contribute to a Sommerfeld enhancement. This enhancement arises from non-relativistic multiple exchanges of the light mediator, which increase the annihilation rate at low velocities~\cite{Hisano:2006nn, Cirelli:2007xd}. We use the analytic approximation given in Refs.~\cite{Slatyer:2009vg,Feng:2010zp, Cassel:2009wt}. We incorporate the Sommerfeld enhancement by multiplying the annihilation cross-section appearing in~\cref{eq:cosmic_sigv} and~\cref{eq:gal_sigv} by the enhancement factor, $ \sigma  \rightarrow \sigma   S(v_{\rm rel})$. This correction is applied both to the present-day relevant for the TRIDENT or H.E.S.S. sensitivity and to the thermally averaged annihilation cross-section at freeze-out. Due to both the t-channel and Sommerfeld enhancement, it is not trivial to see exactly how the limits vary for different couplings. When the s-channel annihilation dominates, $\langle\sigma v\rangle \sim (g_{Z^\prime}^{\rm SM} g_{Z^\prime}^{\rm DM})^2/(4m_\chi^2-m_{Z^\prime}^2)^2$, and smaller couplings simply require smaller masses accordingly.

In addition to the TRIDENT projections we show the constraints on this model from the most sensitive indirect and direct detection results in shaded purple and blue, respectively. The charged lepton decays, $\tau^+\tau^{-}$ and $\mu^+\mu^-$ occurring at tree-level dominate the $\gamma$-ray limits for this model~\cite{HESS:2022ygk}. By contrast, the direct-detection signal is produced only via the loop-induced kinetic mixing, Eq.~(\ref{eq:epsilon}). To determine the elastic nucleon cross-section, we follow Refs.~\cite{Evans:2017kti,Bauer:2018onh,Pan:2018dmu,Foldenauer:2019vgn,Alonso-Gonzalez:2025xqg} and determine that when $m_{Z'} > 500 \, \rm GeV$, the DM-neutron cross-section is zero at first order and the DM-proton cross-section can be written as
\begin{equation}
    \sigma_{\chi p}=\frac{m_\chi^2 m_p^2}{ \pi (m_\chi+m_p)^2} \frac{e^2 (g_{Z^\prime}^{\rm DM})^2 \epsilon^2}{m_{Z'}^4}\, .
\end{equation}

Furthermore, we also show by way of a green dashed line, where the irreducible neutrino background from coherent elastic neutrino-nucleus scattering will come into effect~\cite{Billard:2013qya,O_Hare_2021}. At these energies the dominant neutrino source is atmospheric neutrinos which will be indistinguishable from a DM particle with mass above $\sim 50\,{\rm GeV}$.  We can see that the TRIDENT sensitivities calculated in this work reach deep into the neutrino floor.

For the thermal relic and the TRIDENT constraint, the lines in the left panel of Fig.~\ref{fig:Lmu_Ltau_model} would be unchanged for the $L_e-L_\mu$ and $L_e-L_\tau$ models. However, the PandaX and neutrino floor lines would move a little to the right due to a slight increase in the kinetic mixing. For the H.E.S.S. constraint, a slight weakening of the limit would occur for the $L_e-L_\mu$ model because the $e^+e^-$ and $\mu^+\mu^-$ final states are less constrained with respect to $\tau^+\tau^-$. Furthermore, the $L_e-L_{\mu/\tau}$ models are under much stronger constraints from LEP-2~\cite{Electroweak:2003ram}. In particular, above $m_{Z^\prime}\gtrsim209\,{\rm GeV}$, the coupling is constrained to be $g_{Z^\prime}\le 0.044 \,m_{Z^\prime}/(200\,{\rm GeV})$~\cite{Buckley:2011vc,Dasgupta:2023zrh}, which would provide a constraint on the left panel of Fig.~\ref{fig:Lmu_Ltau_model} of $m_{Z^\prime}\geq 4.5 \, {\rm TeV}$. \\

In the right panel of Fig.~\ref{fig:Lmu_Ltau_model} we focus on the scenario where neutrino telescopes give the only relevant constraints. This is the secluded U(1)$_{L_\mu-L_\tau}$ model and occurs when the $Z^\prime$ mediator is very light, and $g_{Z^\prime}^{\rm DM}\gg g_{Z^\prime}^{\rm SM}$~\cite{Pospelov:2007mp}. In this case the relic abundance can still be achieved via dark sector freeze-out. The most interesting possibility for TRIDENT is when $m_{Z^\prime}<2\,m_\mu$ which results in ${\rm Br}_{4\nu}\approx1$. Light bosons are comprehensively searched for by fixed target and low energy facilities, such as NA64~\cite{NA64:2024klw} and Muon g-2~\cite{Muong-2:2025xyk}. For $m_{Z^\prime}\leq 2m_\mu$, both these experiments require $g^{\rm SM}_{Z^\prime}\lesssim5\times10^{-4}$. Experiments such as LDMX~\cite{Kahn:2018cqs}, a future muon beam-dump experiment~\cite{Cesarotti:2023sje} and astrophysical observations of white dwarf cooling~\cite{Foldenauer:2024cdp} have the potential to constrain this region further. Finally, constraints from cosmological probes require that $m_{Z^\prime}\gtrsim 3\,{\rm MeV}$~\cite{Escudero:2019gzq}.

Once we embed dark matter into this model, the direct detection constraints are much stronger than the pure $Z^\prime$ searches. This typically requires $g^{\rm SM}_{Z^\prime}\lesssim 10^{-6}$ for $m_{\chi}\sim 1\,{\rm TeV}$. The lower bound on $g^{\rm SM}_{Z^\prime}$ comes from Big Bang Nucleosynthesis, requiring the $Z^\prime$ lifetime to be $\lesssim 1\,{\rm s}$, which in turn requires $g^{\rm SM}_{Z^\prime}\gtrsim 10^{-10}$ between $3-200\,{\rm MeV}$. Even still, the annihilation is not entirely invisible to other searches such as the CMB~\cite{Slatyer:2015jla} or even $\gamma$-ray searches through electroweak corrections or the loop-suppressed $Z^\prime$ decay to electrons.

For the entire range of $m_{\chi}$ in the right panel of Fig.~\ref{fig:Lmu_Ltau_model} the Sommerfeld enhancement is in effect, allowing our predicted sensitivity to overcome the uncertainty on the DM halo. The solid red line is for the NFW profile and the red shaded region brackets the uncertainty in the halo as described in Section~\ref{subsec:dmhalo}. We see that when $m_{\chi}\gtrsim 600\,{\rm GeV}$, TRIDENT's sensitivities will be able to confidently exclude this scenario for thermally produced DM.   

We quickly comment on other proposed experimental tests for this model, many of which are reviewed in Ref.~\cite{Bernal:2025szh}. The key limitation for collider searches is the achieved centre-of-mass energies, the largest of which is the LHC, but since our model does not couple to quarks at tree-level, most limits are not relevant.  In fact, measurements at LEP-II remain the most stringent, $m_{Z^{\prime}}\gtrsim 209\,{\rm GeV}$~\cite{OPAL:2003kcu, ALEPH:2006jhv}. This underscores the power a future lepton collider could have on this model~\cite{Huang:2021nkl, InternationalMuonCollider:2025sys, CEPCStudyGroup:2018ghi}.

\section{Summary and Outlook}
\label{sec:summary}

In this paper we have presented the first DM annihilation sensitivity forecasts for TRIDENT. 
%\sout{This is an important step for establishing the analysis framework and can be used to inform any pre-deployment optimisation.}
Our results show that cascade events will bring substantial improvements to the DM sensitivity, by over a factor of $2$ compared to a track-only analysis. This demonstrates the significant impact precise reconstruction of cascade events can bring, further informing ongoing pre-deployment detector optimisation studies. In fact, the cascade sensitivity result shows that TRIDENT will be able to probe the thermal relic cross-section values for DM masses between $10^3-10^5\,{\rm GeV}$. We obtain sensitivities down to $\langle\sigma v\rangle\approx5\times10^{-27}\,{\rm cm}^3\,{\rm s}^{-1}$ at $m_\chi=10\,{\rm TeV}$. This area of parameter space is still untested and is one of the last regions for the `WIMP miracle'. 

As always with indirect detection, one has to contend with uncertainties in the DM distribution, which will weaken TRIDENT's sensitivity. Additionally, the background flux of neutrinos originating from the interactions of hadronic cosmic rays and interstellar gas remains largely unknown and can limit the sensitivity further. 
We estimate the effect this background will have on neutrino telescopes' search for DM annihilation signals from the Galactic Centre. We believe that this is the first time such a background has been incorporated and we find that its impact is greater at energies above $\sim 10\,{\rm TeV}$ where at most the sensitivities are weakened by a factor of $\sim 2$. 
We also highlight the possibility that this background could mimic a DM signal, motivating further research and the development of analysis techniques that distinguish between the two signals.  

In order to emphasise the future impact of TRIDENT we contextualise our model-independent limits with a specific particle physics model, the U(1)$_{L_i-L_j}$, which represents a more realistic scenario than $100\%$ annihilations into neutrinos. We demonstrate that TRIDENT remains sensitive to parameter values consistent with a thermal relic, as annihilations into neutrinos are still substantial. At energies above $1\,{\rm TeV}$, $\gamma$-ray telescopes and direct dark matter detection searches lose sensitivity, opening a complementary discovery window for neutrino telescopes. Due to its large exposure to the Galactic Centre and excellent angular resolution, TRIDENT is uniquely positioned to exploit this window. For example, when $m_{Z^\prime}$ and $m_\chi$ are both at $\sim 10\, {\rm TeV}$ scale, TRIDENT will probe deep into the `neutrino fog', where direct detection experiments will struggle to penetrate. There remain substantial regions where both direct and indirect detection experiments complement each other and would provide independent verification of any signal, which may be needed given the Galactic plane neutrino background. The accompanying $\gamma$-ray signals predicted by the model will also be a source of complementarity aiding both background and particle model discrimination. Furthermore, since relic DM above $1\,{\rm TeV}$ requires resonances or leads to Sommerfeld enhancements, the velocity distribution in the DM profile could be more important than usually assumed for velocity independent annihilation. The TRIDENT sensitivity we report here motivates a more careful analysis.

\acknowledgments
The authors would like to thank the following individuals for their helpful comments and discussions: C.~Arina, P.~Foldenauer, S.~Hernández~Cadena, X.~G.~He, Q.~R.~Liu, X.~G.~Lu, S.~K.~Mondal, K.~Ng, H.~B.~Shao, S.~H.~Hao, and T.~L.~Zhu. S. Kaci acknowledges the support of G.~Giacinti, funded by the National Natural Science Foundation of China under Grant Nos. 12350610239, and 12393853. 
A. Cheek acknowledges the support of S. Ge, funded by the National Natural Science Foundation of China (Grant Nos. 12425506, 12375101, 12090060, and 12090064) and the Double First-Class startup funds from Shanghai Jiao Tong University. 
X. Xiang would like to thank the Double First-Class startup funds by Shanghai Jiao Tong University. 
I. Morton-Blake also acknowledges the National Natural Science Foundation of China Grant No. 12350410499, the Shanghai Science and Technology Commission Grant No. 24590780100 and the Kuan-Cheng Wang Education Foundation.

\clearpage

\appendix

\section{Galactic Plane Neutrino Background: Additional channels}
\label{app:GP}

We show the impact of the Galactic plane neutrino background for the $b\bar{b}$, $\tau^+\tau^-$ and $4\nu$ channels in Fig.~\ref{fig:gp_appendix}, following the approach of Fig.~\ref{fig:galactic_plane} and~\cref{subsec:gp_bkg}. The background has a smaller effect on tracks owing to their better angular resolution. For channels with distinct spectral features, such as the $\nu\bar{\nu}$ $\delta$-peak and the $4\nu$ box-shape, the high energy resolution for cascades mitigates the influence of the Galactic plane background. As with Fig.~\ref{fig:galactic_plane} the central limit degrades above $\sim 1\,{\rm TeV}$ with the effect saturating around $10\,{\rm TeV}$.

\setcounter{figure}{0}
\renewcommand{\thefigure}{A\arabic{figure}}

\begin{figure}[h!]
  \centering
  \begin{subfigure}[t]{.4\textwidth}
    \centering
    \includegraphics[width=\linewidth]{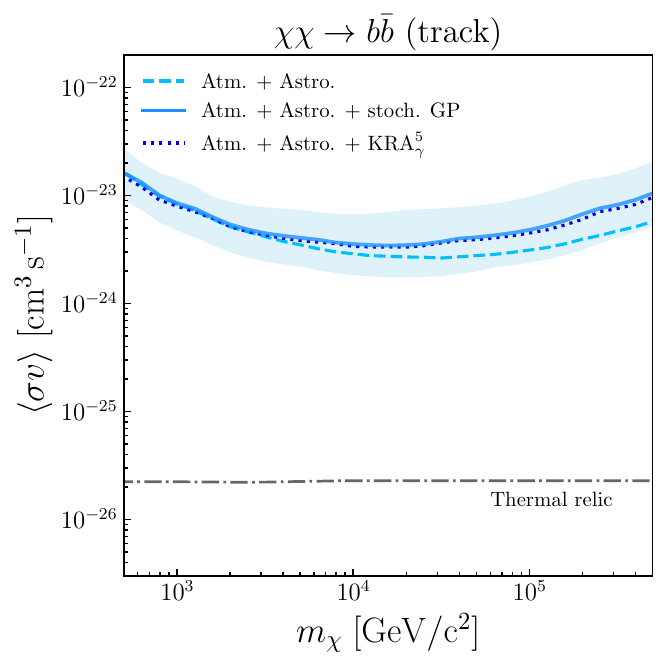}
    % \caption{Track, $\tau^+\tau^-$ ($\psi<5^\circ$)}
    % \label{fig:numu_gp_bb}
  \end{subfigure}\hfill
  \begin{subfigure}[t]{.4\textwidth}
    \centering
    \includegraphics[width=\linewidth]{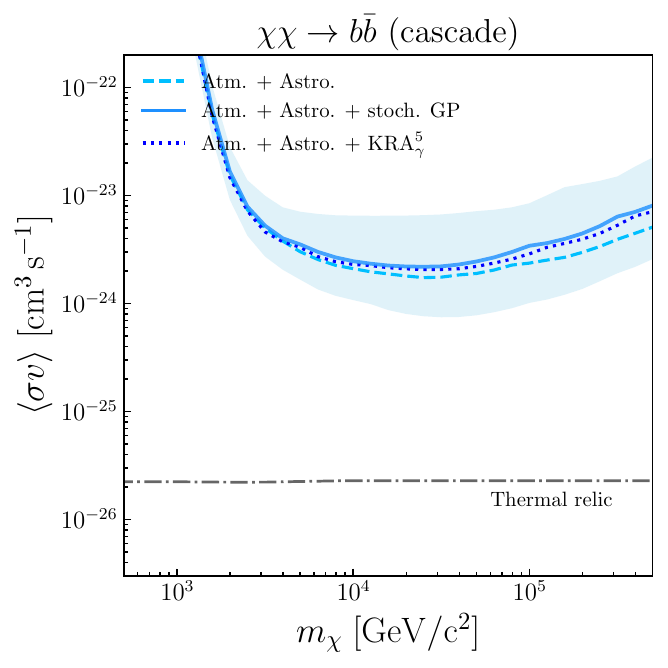}
    % \caption{Cascade, $\tau^+\tau^-$ ($\psi<20^\circ$)}
    % \label{fig:cascade_gp_bb}
  \end{subfigure}
  
  % \vspace{0.3em} 
  
  \begin{subfigure}[t]{.4\textwidth}
    \centering
    \includegraphics[width=\linewidth]{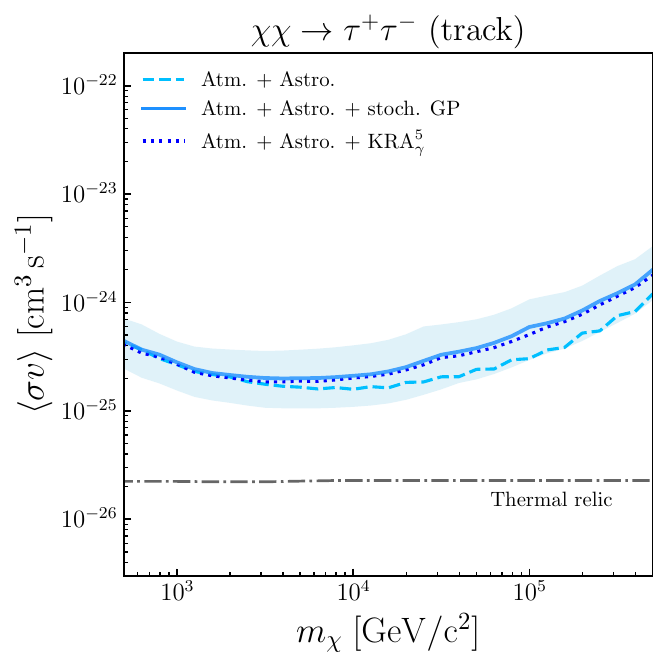}
    % \caption{Track, $\nu\bar{\nu}$ ($\psi<5^\circ$)}
    % \label{fig:numu_gp_tautau}
  \end{subfigure}\hfill
  \begin{subfigure}[t]{.4\textwidth}
    \centering
    \includegraphics[width=\linewidth]{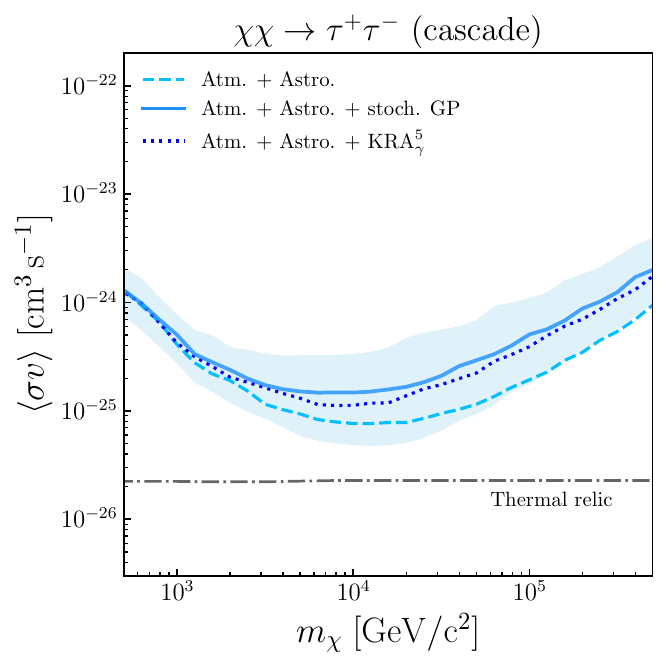}
    % \caption{Cascade, $\nu\bar{\nu}$ ($\psi<20^\circ$)}
    % \label{fig:cascade_gp_tautau}
  \end{subfigure}

  \begin{subfigure}[t]{.4\textwidth}
    \centering
    \includegraphics[width=\linewidth]{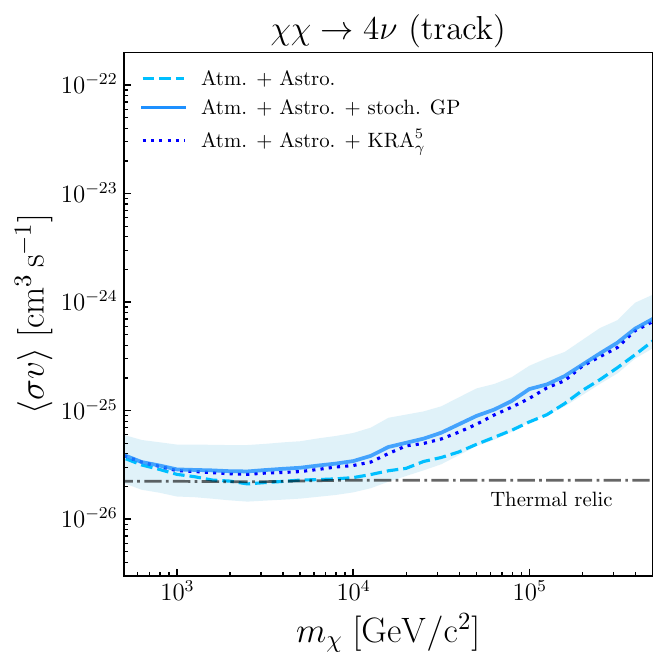}
    % \caption{Track, $\nu\bar{\nu}$ ($\psi<5^\circ$)}
    % \label{fig:numu_gp_4nu}
  \end{subfigure}\hfill
  \begin{subfigure}[t]{.4\textwidth}
    \centering
    \includegraphics[width=\linewidth]{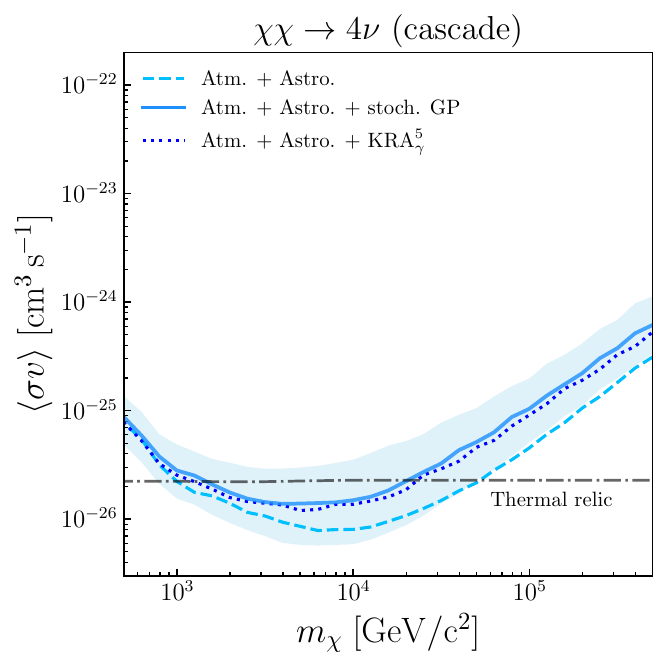}
    % \caption{Cascade, $\nu\bar{\nu}$ ($\psi<20^\circ$)}
    % \label{fig:cascade_gp_4nu}
  \end{subfigure}

  \caption{DM signal sensitivity in the $b\bar b$, $\tau^+\tau^-$ and $4\nu$ annihilation channels, shown from top to bottom, respectively, for track (left) and cascade (right) events. Results are shown without the Galactic-plane background (dashed lines) and with its impact modelled using stochastic Galactic-plane maps (solid lines) or the $\mathrm{KRA}^5_\gamma$ template (dotted lines). We assume 10 years of full detector exposure. Coloured bands represent the 68\% uncertainty intervals associated with the stochastic Galactic-plane realisations.} 
  \label{fig:gp_appendix}
\end{figure}

%\clearpage

\bibliographystyle{JHEP}
\bibliography{ref}

\end{document}